\documentclass[letterpaper]{JHEP3}
\let\ifpdf\relax
\usepackage{
ifpdf,
latexsym,
amsmath,
amsfonts,
amssymb,
enumerate,
graphicx,
subfigure
}

\usepackage[latin1]{inputenc}
\pdfoutput=1
\usepackage{euscript,mathrsfs}
\usepackage{bbm}

\newcommand{\be}{\begin{equation}}
\newcommand{\ee}{\end{equation}}

\newcommand{\ba}{\begin{eqnarray}}
\newcommand{\ea}{\end{eqnarray}}
\newcommand{\nn}{\nonumber}
\newcommand{\ra}{\rightarrow}
\newcommand{\half}{\frac{1}{2}}

\DeclareMathOperator{\area}{area}

\title{Strong subadditivity and the covariant holographic entanglement entropy formula}

\author{
Robert Callan, Jianyang He, and Matthew Headrick \\
Martin Fisher School of Physics, Brandeis University, Waltham, Massachusetts, USA
}

\abstract{
Headrick and Takayanagi showed that the Ryu-Takayanagi holographic entanglement entropy formula generally obeys the strong subadditivity (SSA) inequality, a fundamental property of entropy. However, the Ryu-Takayanagi formula only applies when the bulk spacetime is static. It is not known whether the covariant generalization proposed by Hubeny, Rangamani, and Takayanagi (HRT) also obeys SSA. We investigate this question in three-dimensional AdS-Vaidya spacetimes, finding that SSA is obeyed as long as the bulk spacetime satisfies the null energy condition. This provides strong support for the validity of the HRT formula.
}

\preprint{BRX-TH-645}

\begin{document}

\section{Introduction}

Strong subadditivity (SSA) is a fundamental property of entropy and a cornerstone of quantum information theory. It states that, given a tripartite quantum system $ABC$ and a joint density matrix $\rho(ABC)$, the entropies of the subsystems obey the following inequalities:
\begin{eqnarray}
S(AB)+S(BC)-S(ABC)-S(B)&\ge& 0 \label{SSA1}\\
S(AB)+S(BC)-S(A)-S(C) &\ge&0\,.\label{SSA2}
\end{eqnarray}
SSA is a general theorem that depends only on basic facts about Hilbert spaces and the definition of the von Neumann entropy. Several proofs of it exist but none is very simple. An elementary consequence of SSA is that the mutual information $I(A:B):=S(A)+S(B)-S(AB)$, which measures the total amount of correlation between the subsystems, increases when we adjoin another subsystem to either argument: $I(A:BC) \ge I(A:B)$. As a monotonicity property---in space, rather than time---SSA is thus in some sense analogous to the second law of thermodynamics, and it plays a similarly fundamental role in quantum information theory to that played by the second law in statistical mechanics.

It was shown by Headrick and Takayanagi \cite{Headrick:2007km} that Ryu and Takayanagi's (RT) conjectured holographic entanglement entropy formula \cite{Ryu:2006bv,Ryu:2006ef} generally obeys the SSA property. (See Section 2 for a brief review of the RT formula and this proof.) As no proof or derivation of the RT formula exists, the fact that it obeys this highly non-trivial property constitutes a crucial check on its validity. The proof is remarkably simple---far simpler than any of the proofs of SSA itself. The fact that the RT formula obeys SSA is analogous to the black-hole area theorem (the event horizon area increases with time), which was a crucial motivation for Bekenstein and Hawking's identification of the horizon area with the black hole's thermodynamic entropy.

The RT formula is restricted to states represented holographically by static spacetimes and to regions of the boundary lying inside constant-time surfaces. Hubeny, Rangamani, and Takayanagi (HRT) \cite{Hubeny:2007xt} subsequently proposed a covariant generalization, in which the entanglement entropy is given by the area of a certain extremal codimension-2 spacelike surface (reviewed in Section 2). While the HRT formula has been applied to a variety of time-dependent systems, it has so far been subjected to somewhat less stringent testing than the original RT formula. In particular, Headrick and Takayanagi's proof of SSA for RT cannot be straightforwardly generalized to HRT, and it is far from obvious whether or why the areas of extremal spacelike surfaces obey these global inequalities. Deciding whether HRT obeys SSA is thus an important question, for two reasons: First, it is a strong test of that very useful formula. Second, if true, the statement that HRT obeys SSA would be a very interesting new theorem in general relativity, which---much like the black-hole area theorem---would give new insight into the relationship between areas and entropies in quantum gravity.

In this paper we begin to address the question of whether HRT obeys SSA by a study of examples. We restrict ourselves to the simplest non-trivial context for field-theory entanglement entropies, namely single intervals in two-dimensional field theories. (In order that not only $A,B,C$ be single intervals but also their unions $AB,BC,ABC$ appearing in \eqref{SSA1}, \eqref{SSA2}, we take them to be adjacent and in the order $A$--$B$--$C$.) According to the HRT formula, the entropy of such an interval is simply given by the length of the shortest spacelike geodesic in the three-dimensional bulk connecting its endpoints (subject to a certain homology condition \cite{Fursaev:2006ih}).\footnote{Because of the homology condition, in certain cases the extremal surface may also include other connected components that do not extend to the boundary (for example that wrap the horizon of a black hole). In the present investigation, for simplicity we will neglect this possibility.} Despite its relative simplicity, this context allows us to address the essential issues that arise in going from the RT to the HRT formula, including having a non-static bulk and having boundary regions that do not lie on a constant-time surface. More specifically, we consider planar thin-shell AdS${}_3$-Vaidya spacetimes, as these are the simplest time-dependent asymptotically AdS${}_3$ exact solutions to the Einstein equation. In particular, their simplicity allows us to solve the geodesic equation exactly (using the method of \cite{Balasubramanian:2011ur}), and systematically explore the space of spacelike geodesics.\footnote{The same investigation was also carried out for thick-shell Vaidya spacetimes, yielding substantially identical results. Here we report the thin-shell results, since those were obtained analytically whereas the thick-shell case required numerical integration of the geodesic equation.}${}^,$\footnote{Several papers, including \cite{AbajoArrastia:2010yt,Albash:2010mv,Balasubramanian:2010ce,Balasubramanian:2011ur,Allais:2011ys,Aparicio:2011zy,Balasubramanian:2011at}, have previously considered entanglement entropies in AdS$_3$-Vaidya spacetimes (mainly focusing on the case of constant-time intervals).} In order to investigate the role of the null energy condition (NEC), we consider both positive- and negative-energy infalling matter. In Section 3, we study intervals that lie on a constant-time surface of the boundary (with respect to the Lorentz frame distinguished by the system's spatial translation symmetry). In Section 4, we turn to more general spacelike intervals. This is an important case to consider when hunting for violations of SSA, because, as we explain there, certain such configurations saturate SSA in the static limit.

Our investigation of Vaidya found no violations for positive-energy infalling matter. If violations of SSA were generic, we would expect them to appear in the Vaidya case, so these results provide a highly non-trivial check on the HRT formula. On the other hand, we observed that negative-energy matter leads to violations, which indicates that, if the HRT formula is correct, the NEC somehow plays a crucial role in maintaining its consistency. (The correlation between SSA and the NEC was also noted by Allais and Tonni \cite{Allais:2011ys}, based on the study of constant-time intervals.) The NEC plays a key role in assuring the consistency of the area-entropy relationship in other gravitational contexts, including the area theorem and the covariant holographic entropy bound \cite{Bousso:1999xy,Bousso:2002ju}, so its appearance here is very natural.\footnote{Recent work on holographic c-theorems based on entanglement \cite{Liu:2012ee,Myers:2012ed} also indicated that the NEC is a necessary condition.}

While our positive results for AdS${}_3$-Vaidya are a good start, there are clearly many more cases that should be checked in order to be confident that HRT obeys SSA generally. An obvious generalization would be to higher dimensions. Another would be to more complicated shapes and configurations of boundary regions. A third would be the case where the bulk has non-trivial topology, and the homology condition comes into play. Fourth, there are many cases (with and without non-trivial bulk topology) where there is a phase transition and the minimal extremal surface switches from one topology to another as the boundary region is varied. Finally, in order to better understand the role of the NEC, one could investigate in examples whether there is a direct connection between the violation of SSA and the violation of holographic $c$-theorems that are seen when the NEC is violated \cite{Liu:2012ee,Myers:2012ed}.

Of course, ultimately one does not want merely a laundry list of examples where SSA is satisfied (assuming it is indeed satisfied). Rather, one wants to understand how the HRT formula pulls off such a trick: one wants a proof, perhaps along the lines of the proof of the area theorem. As argued above, such a proof would not only provide very strong support to the HRT formula, but would almost certainly provide new, deep insight into the connection between areas and entropies in gravitational theories, a connection about which we have so many hints yet still so little real understanding.

\section{Review: Holographic entanglement entropy formulas}

\subsection{RT}

Let us briefly recall the Ryu-Takayanagi (RT) formula \cite{Ryu:2006bv,Ryu:2006ef}, and the proof that it obeys SSA \cite{Headrick:2007km}. The formula applies to field theories that are holographically dual to Einstein gravity, in states that are described by static classical geometries. Staticity implies the existence of distinguished constant-time slices, both in the bulk and on the boundary; the RT formula gives the entanglement entropy of subregions of such a boundary slice, and we work entirely within the corresponding bulk slice, with its Euclidean induced metric---the time direction and components of the metric play no role. Specifically, given a boundary region $A$, the RT formula says that its entanglement entropy is given by the area of the minimal surface in the bulk within a certain topological class:\footnote{The area is evaluated with respect to the Einstein-frame metric. No other fields enter into the formula.}
\begin{equation}\label{RT}
S(A) = \frac1{4G_{\rm N}}\min_{m\sim A}\area(m)\,;
\end{equation}
$m\sim A$ means that $m$ and $A$ are homologous, i.e.\ there exists a bulk region $r$ such that $\partial r = m\cup A$ (and therefore necessarily $\partial m = \partial A$, i.e.\ $m$ is ``anchored" on the boundary along $\partial A$).

\FIGURE{
\includegraphics[width=3in]{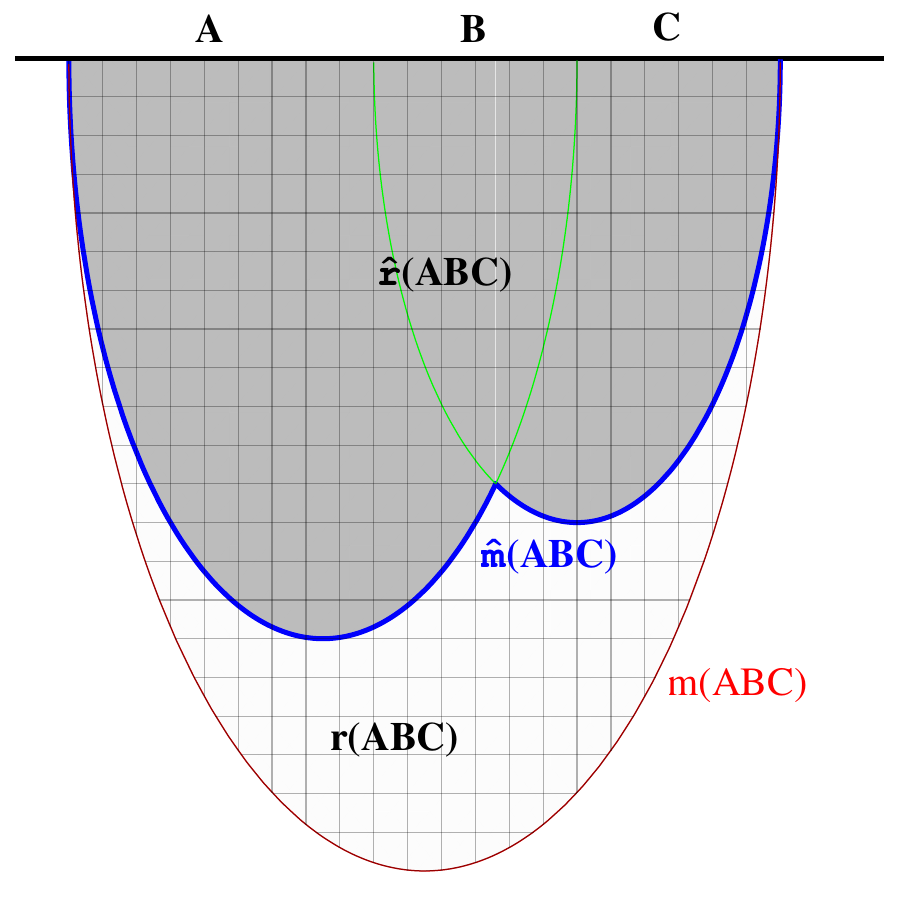}
\includegraphics[width=3in]{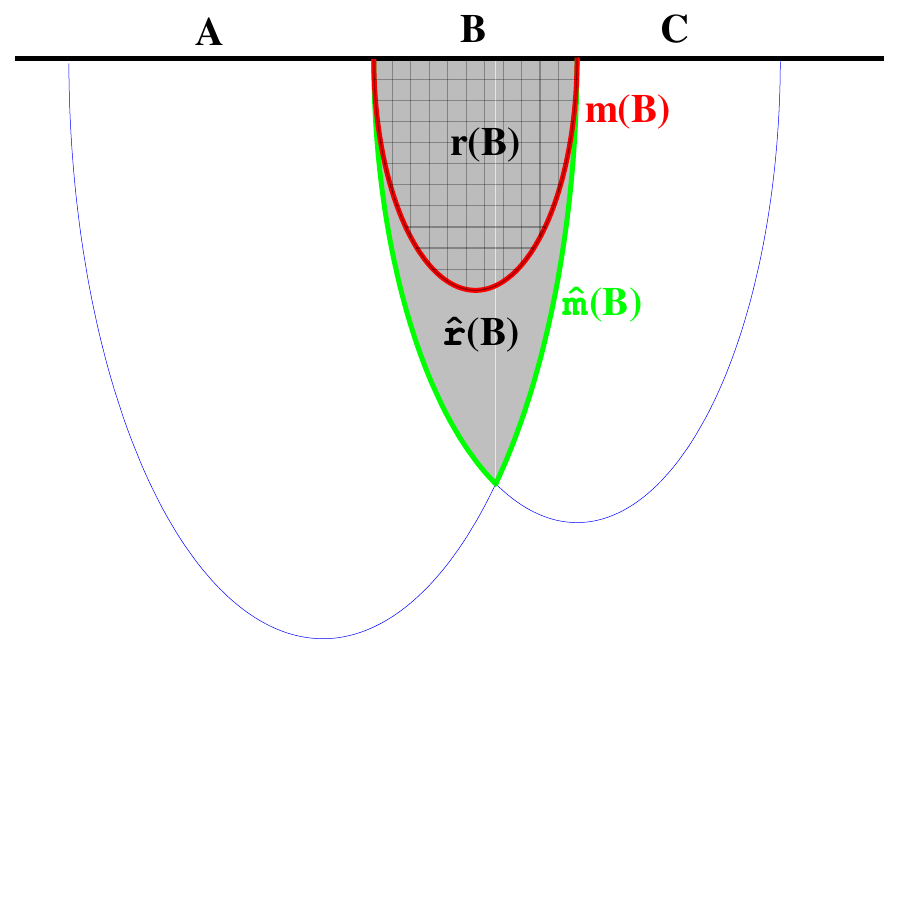}
\caption{Sketch of the different regions and surfaces appearing in the holographic proof of SSA. $A,B,C$ are boundary regions. Each region, as well as their unions $AB$, $BC$, and $ABC$, has a corresponding bulk region $r(AB)$ etc. The bulk region $\hat r(ABC)$ is defined as $r(AB)\cup r(BC)$ (left diagram), while $\hat r(B) = r(AB)\cap r(BC)$ (right diagram). Equations \eqref{step1}, \eqref{step2}, and \eqref{step3p} are hopefully clear from the diagrams.}
\label{proofcartoon}
}

The proof that \eqref{RT} obeys \eqref{SSA1} is as follows (see Fig.\ \ref{proofcartoon}) \cite{Headrick:2007km}. Denote the minimizing surface and region $m(A)$ and $r(A)$ respectively. We now consider the region $\hat r(ABC):=r(AB)\cup r(BC)$, and decompose its boundary into the part lying along the boundary of the spacetime, which is $AB\cup BC = ABC$, and the part lying in the bulk, which we denote $\hat m(ABC)$: $\partial\hat r(ABC) = ABC\cup\hat m(ABC)$, i.e.\ $\hat m(ABC)\sim ABC$. Hence
\begin{equation}\label{step1}
S(ABC) = \frac1{4G_{\rm N}}\area(m(ABC)) \le \frac1{4G_{\rm N}}\area(\hat m(ABC))\,.
\end{equation}
Similarly defining $\hat m(B)$ as the part of $\partial(r(AB)\cap r(BC))$ lying in the bulk, we have
\begin{equation}\label{step2}
S(B) = \frac1{4G_{\rm N}}\area(m(B)) \le \frac1{4G_{\rm N}}\area(\hat m(B))\,.
\end{equation}
The last step is to note that
\begin{equation}\label{step3p}
\hat m(ABC)\cup\hat m(B) = m(AB)\cup m(BC)
\end{equation}
(this should be clear from the sketch in Fig.\ \ref{proofcartoon}, but can easily be shown rigorously by decomposing $m(AB)$ into the part lying inside $r(BC)$ and the part lying outside it, and similarly with $m(BC)$), so that
\begin{equation}\label{step3}
\begin{split}
\frac1{4G_{\rm N}}\area(\hat m(ABC))+\frac1{4G_{\rm N}}\area(\hat m(B)) &= \frac1{4G_{\rm N}}\area(m(AB))+\frac1{4G_{\rm N}}\area(m(BC)) \\ &= S(AB)+S(BC)\,.
\end{split}
\end{equation}
Combining \eqref{step1}, \eqref{step2}, and \eqref{step3}, we arrive at \eqref{SSA1}. A similar argument leads to \eqref{SSA2}.\footnote{It was recently shown \cite{Hayden:2011ag} using a similar method that the RT formula also implies the inequality
\begin{equation}
S(AB)+S(BC)+S(AC)-S(ABC)-S(A)-S(B)-S(C)\ge0\,.
\end{equation}
Unlike SSA, however, this inequality is not obeyed by general quantum systems, but rather is a special property of holographic theories.}

This proof uses precisely the input that is given by the RT formula, namely the homology and minimal-area conditions. Nothing more needs to be assumed about the topology or geometry of the bulk spacetime. (Not even the null energy condition needs to be imposed.) Such generality is both a blessing and a curse, as no constraints on the spacetime can be deduced by requiring that SSA be satisfied. In fact, even the minimal-area condition can be relaxed a bit, as the proof goes through if we replace the area functional in \eqref{RT} with any extensive functional of $m$ \cite{Headrick:2007km}. Indeed, it is believed that, in the presence of higher-derivative corrections to the classical bulk gravitational action, such as those arising in string theory, the area functional gets corrected \cite{Fursaev:2006ih,Headrick:2010zt} (although the form of the latter corrections is known only in certain cases \cite{deBoer:2011wk,Hung:2011xb}); if this is true, then SSA is automatically safe, and we learn nothing about the form of the corrections (aside from their extensivity). The same does not apply to bulk quantum corrections, whose effect on the entanglement entropy is not simply to correct the functional being minimized in \eqref{RT}, as can be argued for example by looking at the mutual information between distant regions \cite{Headrick:2010zt}; otherwise little is known about these corrections, whose form may be usefully constrained by SSA.

\subsection{HRT}

It is clear that a major limitation of the RT formula is the restriction to static bulk spacetimes and to boundary regions sitting inside constant-time slices. If we search for a covariant generalization, then we can no longer appeal to the constant-time slice in the bulk, with its Euclidean induced geometry, but rather have to contend with the full Lorentzian spacetime. In a Lorentzian spacetime, the area of a spacelike surface with fixed boundary conditions can be made arbitrarily close to zero by adding wiggles in the time direction, so the minimal-area prescription does not make sense and the RT formula must be altered. Motivated in part by Bousso's covariant entropy bound \cite{Bousso:1999xy,Bousso:2002ju}, Hubeny, Rangamani, and Takayanagi (HRT) \cite{Hubeny:2007xt} suggested replacing the minimal surface with an extremal one, i.e.\ a spacelike surface that extremizes the area functional, or equivalently has vanishing mean curvature; in case there are multiple extremal surfaces, we are instructed to choose the one with the smallest area. So we have
\begin{equation}\label{HRT}
S(A) = \frac1{4G_{\rm N}}\min_{\text{extremal }m\sim A}\area(m)\,.
\end{equation}
(As in the RT formula, $m$ has the same dimensionality as $A$, so it is codimension two.) The choice to replace a minimization procedure by an extremization one when passing from a static situation to a dynamical one is a very natural one; for example in classical mechanics we go from minimizing the energy for a static solution to extremizing the action for a time-dependent one. Similarly, while there is no path-integral derivation of either the RT or HRT formula, the HRT formula is reminiscent of the prescription of Euclidean quantum gravity; there, since the Euclidean Einstein-Hilbert action is unbounded below, rather than minimizing it, one looks for the extremum with the smallest action.

The HRT formula passes several basic consistency checks. For example, it correctly assigns the same entropy to any two regions with the same (boundary) causal domain.\footnote{In the field theory, this property follows from the fact that the respective reduced density matrices $\rho(A),\rho(A')$ are related by a unitary transformation. Using the HRT formula, it can be argued as follows. The region $A$ is clearly homologous to the future boundary of its causal domain. $A'$ is as well, and since they have the same causal domain, $A\sim A'$. Hence in \eqref{HRT} the right-hand side is minimized over the same set of surfaces $m$.} It also correctly reproduces the entanglement entropy in a few simple examples where the latter is known, such as for a two-dimensional CFT in at finite temperature \cite{Hubeny:2007xt}. Nonetheless, it is fair to say that the HRT formula has so far been subjected to far fewer tests than the RT formula. In particular, a crucial consistency test is whether it obeys SSA. Unfortunately, the proof of SSA given above does not admit a straightforward generalization to the HRT formula. Firstly, because the regions $r(AB),r(BC)$ are not restricted to a given constant-time slice, one cannot necessarily take their union and intersection to obtain sensible spacelike regions. Secondly, $m(B)$ and $m(ABC)$ are not minimal surfaces, but merely minimal among the extremal surfaces, so even if one somehow defined appropriate surfaces $\hat m(ABC),\hat m(B)$, one would not automatically have \eqref{step1} and \eqref{step2}.

\section{Constant-time intervals}

In this paper we consider a two-dimensional conformal field theory that admits a holographically dual description as classical three-dimensional Einstein gravity. Because we will employ a Vaidya solution, the bulk theory should admit some form of matter whose stress tensor is (at least in some limit) that of null dust; the theory is otherwise unspecified. The field theory will be on Minkowski space (with time and space coordinates denoted $x,t$ respectively), and we will consider two types of bulk solutions: the planar BTZ black hole, and the planar thin-shell AdS${}_3$-Vaidya solution (with infalling matter). The corresponding states (or processes) in the field theory break both its conformal and Lorentz symmetry, but preserve its spatial translation symmetry. The black hole preserves the time-translation symmetry, while the Vaidya solution breaks it.

In general, the entanglement entropy of a spatial region depends only on its causal domain. In two-dimensional Minkowski space, the causal domain of a connected region (i.e.\ a spacelike curve) is determined by its endpoints; hence we need not concern ourselves with the shape of the curve itself. Since we consider states that are translationally invariant in $x$, the entropy is a function of the times $t_1,t_2$ of the endpoints and of their spatial separation $\Delta x$. In this section we will restrict our attention to intervals that sit at constant time, so $t_1=t_2=t$. In subsection 3.1 we review the fact that the entropy of such intervals obeys SSA if and only if it is a concave and monotone-increasing function of $\Delta x$ (see for example \cite{Wehrl:1978zz} for a more detailed discussion). This is a very useful criterion, allowing us to see at a glance in any particular example whether SSA is satisfied, while also helping to build our intuition about the meaning of SSA. In subsection 3.2 we will study the entanglement entropies of constant-time intervals in the Vaidya spacetime, and show that they obey this condition; most of the analysis will be exact, but since the entropy is obtained as a rather complicated implicit function of $\Delta x$, we give a proof-by-plot that the function is indeed increasing and concave. Finally, in order to investigate the possible role of the null energy condition (NEC) in enforcing SSA, in subsection 3.3 we consider the case of a Vaidya spacetime with infalling negative-energy matter, which makes a transition from a black-hole to an empty AdS spacetime. We find indeed that, while the entropy is still an increasing function of $\Delta x$ in this case, it is no longer concave (so \eqref{SSA2} is satisfied, while \eqref{SSA1} is violated). We explain how this result could have been expected by comparing the entropy curves for AdS and BTZ.

\subsection{SSA, concavity, and monotonicity}

Consider a system with one infinite spatial dimension, and denote the coordinate $x$. We work at a fixed time $t$, and suppress the dependence of all entropies on it. If the system has translational symmetry in $x$, then  the entropy of a single interval $A$ depends only on its length, $S(A) = s(l_x)$, where $s:\mathbb{R}^+\to\mathbb{R}$. In this subsection we will show that the entropies of all such intervals obey \eqref{SSA1} if and only if $s$ is concave\footnote{Throughout this paper, by concave we mean weakly concave, i.e.\ for $0<y<1$, $s(yl_1+(1-y)l_2)\ge ys(l_1)+(1-y)s(l_2)$.}, and obey \eqref{SSA2} if and only if it is non-decreasing. As mentioned in the introduction, in order for the regions $AB,BC,ABC$ appearing in \eqref{SSA1},\eqref{SSA2} also to be single intervals, we assume that $A,B,C$ are adjacent and in that order.

Let $a,b,c$ be the lengths of $A,B,C$ respectively (so $S(A)=s(a)$, $S(AB)=s(a+b)$, etc.). Set $y=c/(a+c)$. We have
\begin{equation}
a+b = yb+(1-y)(a+b+c)\,,\qquad b+c = (1-y)b+y(a+b+c)\,.
\end{equation}
If $s$ is concave then, since $0<y<1$,
\begin{equation}
s(a+b)\ge ys(b)+(1-y)s(a+b+c)\,,\qquad s(b+c)\ge(1-y)s(b)+ys(a+b+c)\,.
\end{equation}
Adding these inequalities, we obtain \eqref{SSA1}. On the other hand, if $s$ is not concave then there exist positive numbers $a,b$ such that $s(a+b)<\frac12(s(2a+b)+s(b))$; choosing $c=a$ we obtain $s(a+b)+s(b+c)<s(a+b+c)+s(b)$, violating \eqref{SSA1}.

Similarly, if $s$ in non-decreasing then clearly
\begin{equation}
s(a+b) \ge s(a)\,,\qquad s(b+c)\ge s(c)\,.
\end{equation}
Adding these inequalities we obtain \eqref{SSA2}. Conversely if $s$ is somewhere decreasing then there exist $a,b$ such that $s(a+b)<s(a)$; choosing $c=a$ we arrive at a violation of \eqref{SSA2}.

The simplest examples are provided by conformal field theories at zero and finite temperature. The entropy of an interval is given at zero temperature by \cite{Holzhey:1994we}
\begin{equation}\label{zeroT}
s(l_x) = \frac c3\ln\left(\frac{l_x}\epsilon\right),\qquad
\end{equation}
and at temperature $T>0$ by \cite{Calabrese:2004eu}
\begin{equation}\label{nonzeroT}
s(l_x) = \frac c3\ln\left(\frac1{\pi T\epsilon}\sinh(\pi Tl_x)
\right)
\end{equation}
where $c$ is the theory's central charge and $\epsilon$ is an ultraviolet cutoff length. Both functions are clearly increasing and concave. For theories with holographic duals, the vacuum and thermal states are represented by AdS${}_3$ and the planar BTZ black hole respectively. Since these are static and we are considering constant-time intervals, the RT formula may be applied in order to obtain \eqref{zeroT},\eqref{nonzeroT} \cite{Ryu:2006bv,Ryu:2006ef}, and the SSA property follows from the Headrick-Takayanagi proof. We now turn to the AdS-Vaidya spacetime, where this is not the case.

\subsection{Vaidya spacetime\label{sec:thinshellVaidya}}

The planar AdS${}_3$-Vaidya metric is
\be
\label{eqn:VaidyametricInrv}
ds^2=-(r^2-m(v))dv^2+2drdv+r^2dx^2\,.
\ee
The boundary is at $r\to\infty$, where the $v$ coordinate is identified with the $t$ coordinate. The spacetime satisfies the NEC as long as $m'(v)\ge0$. We will consider two thin-shell solutions, in which the infalling matter is concentrated in a delta-function at $v=0$. In this subsection, we will take
\begin{equation}
m(v) = \begin{cases} 0\,,& v<0 \\ m\,,& v>0 \end{cases}\,,
\end{equation}
where $m>0$, representing a shell of (positive-energy) matter in otherwise empty AdS which collapses to form a BTZ black hole. In subsection 3.3, we will consider the opposite situation,
\begin{equation}
m(v) = \begin{cases} m\,,& v<0 \\ 0\,,& v>0 \end{cases}\,,
\end{equation}
in which a shell of negative-energy matter collapses onto and annihilates a BTZ black hole, leaving behind empty AdS.

In subsubsection 3.2.1, we construct and discuss the Penrose diagram for the positive-energy Vaidya spacetime. This sets the stage for the systematic study in 3.2.2 of the global structure of the spacelike geodesics that are symmetric under $x\to-x$, since those are the ones that give the entanglement entropies of constant-time boundary intervals. We follow the method of \cite{Balasubramanian:2011ur} to solve the geodesic equation in the thin-shell geometry. As noted in \cite{AbajoArrastia:2010yt}, an interesting feature is that some of the geodesics pass behind the horizon; in other words, accordng to HRT the entanglement entropies of some boundary intervals depend on the metric in the region behind the horizon. Finally, we use the solutions to test SSA. Given the theorem proven in subsection 3.1, this amounts to showing that the length of the geodesic is a concave, monotone function of the separation of its endpoints on the boundary. This is shown in Fig.\ \ref{fig:PureSpacelikeIntervalsVaidyavB}. On the other hand, it fails to be concave in the negative-energy case, as shown in Fig.\ \ref{fig:UnphysicalVaidyaLoflxlist}. A heuristic way to understand the reason for this behavior is the following: A geodesic that is anchored at a time $t>0$ will, if the interval is sufficiently short, lie entirely in the AdS region. But if we increase the length of the interval, the geodesic will eventually cross into the BTZ region. So the entanglement entropy $s(l_x)$ will cross over from \eqref{zeroT} to \eqref{nonzeroT}. As \eqref{nonzeroT} has a larger slope than \eqref{zeroT} for the relevant values of $l_x$, concavity fails in the crossover region.

\subsubsection{Penrose diagram}

Before we solve the geodesics, it is helpful to plot the Penrose diagram of Vaidya space, and to have some ideas on the global structure of the spacetime. If the shell is thin, one can consider the two regions, pure $AdS_3$ and BTZ separately, and match them together along the shell later on.

If we define $f(r,v):= r^2-m(v)$, and introduce
\be
v=t+\tilde{r},\hspace{1cm}\mathrm{with~}d\tilde{r}=\frac{dr}{f(r,v)},
\ee
one gets the familiar metric,
\[
ds^2=\left\{
\begin{array}{ll}
-r^2 dt^2+\frac{dr^2}{r^2}+r^2dx^2, & \mathrm{if~} v<0,~{AdS_3} \\
-(r^2-m)dt^2+\frac{dr^2}{r^2-m}+r^2dx^2, & \mathrm{if~} v>0,~\mathrm{BTZ}
\end{array}
\right.
\]
One can easily solve, for AdS${}_3$,
\be
\label{eqn:rtildeAdS}
\tilde{r}=-1/r
\ee
while for BTZ,
\be
\label{eqn:rtildeBTZ}
\tilde{r}=-\frac{1}{\sqrt{m}}\mathrm{arctanh}[\frac{\sqrt{m}}{r}].
\ee
As we will discuss, the mass parameter $m$ is free to set to $1$,
 however if we want to simulate a thick shell spacetime, a variable $m$ would be necessary.
Note that for $x>1$,
\be
\mathrm{arctanh}(x)=\mathrm{arctanh}(1/x)-i\frac{\pi}{2}.
\ee

To find the Penrose diagram of AdS${}_3$, the standard method is to introduce $u=t-\tilde{r}$, and apply a series of coordinate transformations,
\[
\left\{
\begin{array}{ll}
v=t+\tilde{r} \\ u=t-\tilde{r}
\end{array}
\right.
\hspace{1cm}
\left\{
\begin{array}{ll}
v=\ln~V \\ u=-\ln(-U)
\end{array}
\right.
\hspace{1cm}
\left\{
\begin{array}{ll}
V=\tan\tilde{V} \\ U=\tan\tilde{U}
\end{array}
\right.
\]
For BTZ, one should better introduce some coefficients, for example,
\be
v=\frac{1}{a}\ln~V, \hspace{1cm} V=b~\tan\frac{\tilde{V}-\tilde{V}_0}{c}.
\ee
However it turns out only $a$ is necessary, and the real value requires $a=\sqrt{m}$. Thus one has
\[
\left\{
\begin{array}{ll}
v=t+\tilde{r} \\ u=t-\tilde{r}
\end{array}
\right.
\hspace{1cm}
\left\{
\begin{array}{ll}
v=\frac{1}{\sqrt{m}}\ln~V \\ u=-\frac{1}{\sqrt{m}}\ln(-U)
\end{array}
\right.
\hspace{1cm}
\left\{
\begin{array}{ll}
V=\tan\tilde{V} \\ U=\tan\tilde{U}
\end{array}
\right.
\]

\begin{figure}[htbp]
\begin{center}
\includegraphics[scale=.8]{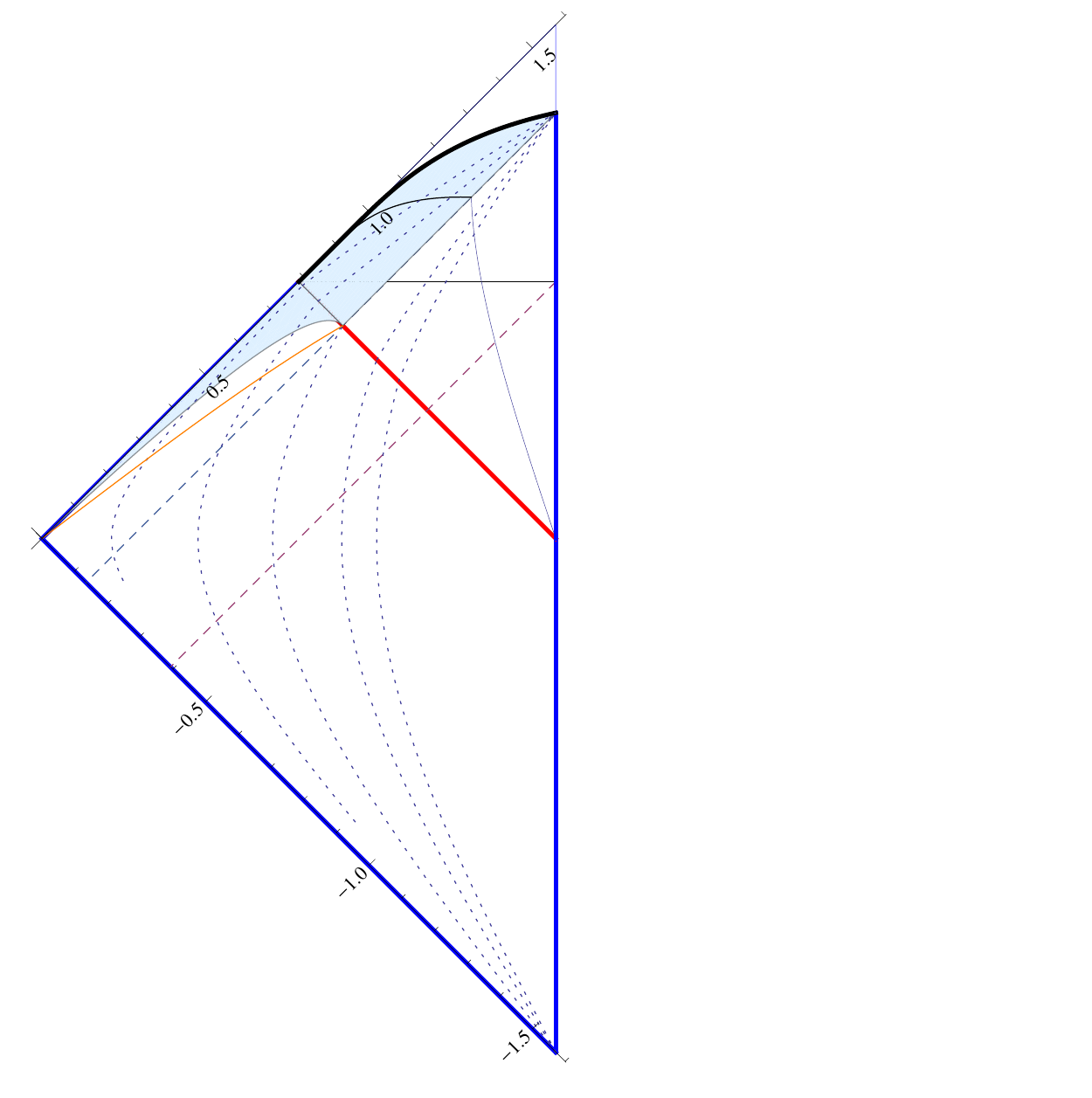}
\caption{Penrose diagram for a thin-shell positive energy Vaidya space. The 45-degree red line is the infalling null shell. The vertical blue line is the asymptotic AdS boundary, and the black curve on the top is $r=0$ singularity. The dashed 45-degree lines are BTZ's horizons and their extension back to AdS region. The dotted curves indicate constant radius tracks. A series of coordinate transformations are imposed on BTZ region, to match the coordinates in AdS along the mass-shell. The asymptotic AdS boundary of BTZ spacetime is intentionally set to be vertical. However the black curved singularity $r=0$ is in general not able to keep horizontal straight simultaneously.
}
\label{fig:Geodesics_Vaidya_ThinShell_Background}
\end{center}
\end{figure}

Since $(r,v)$ are global coordinates, the $(\tilde{V},\tilde{U})$ should be matched at the shell.
With $2\tilde{r}=v-u$, the coordinate matching at $v=0$ shell tells us
\be
u_A=\frac{2}{\sqrt{m}}\tanh\left(\frac{\sqrt{m}}{2}u_B\right).
\ee
The matching in $\tilde{U}$ is,
\be
\ln[\tan(-\tilde{U}_A)]=\frac{2}{\sqrt{m}}\tanh\Big[\frac{1}{2}\ln(-\tan\tilde{U}_B)\Big],
\ee
or
\be
\tilde{U}_B=-\arctan\left[e^{2~\mathrm{arctanh}\left[\frac{\sqrt{m}}{2}\ln(-\tan\tilde{U}_A)\right]}\right].
\ee
The Penrose diagram of a thin-shell Vaidya space is given in Fig.\ref{fig:Geodesics_Vaidya_ThinShell_Background}.\footnote{To mimic a thick-shell Vaidya space, one could impose multiple mass shells in the space, each of which emitted from the asymptotic boundary at different time. The matching of coordinates is much more messy. Keeping both singularity and boundary straight simultaneously is a very tricky work, and in general we believe it is not possible.  Nevertheless it provides us an analytic method for general constructions.} The asymptotic AdS boundary is intentionally kept as vertical, while the behavior of the singularity is more tricky. With the nontrivial matching of coordinates along the shell, it seems that we are not able to find a horizontal straight singularity in Penrose diagram.

\subsubsection{Spacelike geodesics}

To solve the spacelike geodesics, the equations of motion could be found,
\ba
\label{eqn:EOM_Vaidya_randv}
-f(r,v)v'^2+2r'v'+r^2 &=& \frac{r^4}{p_x^2}, \nn\\
r^2-r^2v'^2-rv''+2r'v' &=& 0,
\ea
where the conserved momentum $p_x=r^2\dot{x}$ is introduced. In this note, the superscript prime indicates $\frac{d}{dx}$, and the overdot means $\frac{d}{d\tau}$, the derivative by the affine parameter $\tau$. Since the metric components do not contain $t$ except for along the thin-shell, it would be helpful to introduce another ``conserved" momentum along $t$,
\[
E\equiv f(r,v)\dot{t}=
\left\{
\begin{array}{ll}
r^2\dot{t}, & \mathrm{if~} v<0 \\
(r^2-m)\dot{t}, & \mathrm{if~} v>0
\end{array}
\right.
\]
The system has a scaling invariance under $r_h=\sqrt{m}$,
\be
\{r\ra r_h r,~~t\ra \frac{t}{r_h},~~x\ra \frac{x}{r_h},~~v\ra \frac{v}{r_h},~~p_x\ra r_h p_x,~~E\ra r_h E\}.
\ee
Equivalently one is free to set the mass $m=1$. It makes sense since in Poincare coordinates, there is only one dimensionful scale.

For a spacelike interval on the boundary, time $t=constant$, and due to the symmetry of the spacetime, the geodesic anchored at the endpoints of the interval is symmetric. This simplifies our calculation.

For simplicity, the thin shell is set along $v=0$. On the other hand, since both AdS and BTZ has $\tilde{r}(r\ra\infty)=0$, then the endpoints of geodesic have
\be
v_b\equiv v_{boundary}=(t+\tilde{r})_{boundary}=t_b.
\ee
If $t_b<0$, the whole spacelike geodesic is in AdS bulk, while $t_b>0$, it might cross the shell.

For a spacetime with metric,
\be
ds^2=-f(r)dt^2+\frac{dr^2}{f(r)}+r^2dx^2,
\ee
since there is no $x$-dependence and $t$-dependence in metric, two momenta are conserved,
\be
p_t=-E=g_{tt}p^t=-f(r)\dot{t}, \hspace{1cm} p_x=g_{xx}p^x=r^2\dot{x}.
\ee
The affine parameter condition is
\be
g_{\mu\nu}\frac{\partial x^{\mu}}{\partial \tau}\frac{\partial x^{\nu}}{\partial \tau}=1,
\ee
which leads us
\be
\label{eqn:rdotrprime_AdSBTZ}
\dot{r}^2=f+E^2-\frac{f}{r^2}p_x^2, \hspace{1cm}
(r')^2=\frac{r^4f}{p_x^2}+\frac{r^4}{p_x^2}E^2 -r^2f.
\ee
One can solve the minimum of radius $r$ with $\dot{r}(r_*)=r'(r_*)=0$. For a geodesic with constant time, i.e. $E=0$, one finds $r_*=p_x$.

If the geodesic is totally in $AdS_3$ space, the symmetry requires $E=0$, and
\be
\dot{r}^2=r^2-p_x^2, \hspace{1cm} (r')^2=r^6/p_x^2-r^4, \hspace{1cm} r_*=p_x.
\ee
One can calculate the proper length $L$ and the expansion $l$ along $x$,
\ba
L_{AdS}&=\int d\tau=\int dr/\dot{r}=2\int_{r_*}^{r_{\infty}} \frac{dr}{\sqrt{r^2-p_x^2}}
&=2\log(r+\sqrt{r^2-p_x^2})\Big|_{r_*}^{r_{\infty}}
=2\log(2r_{\infty})-2\log p_x, \nonumber\\
l_{AdS}&=\int dx=\int dr/r' = 2\int_{r_*}^{r_{\infty}} \frac{dr}{r^2\sqrt{\frac{r^2}{p_x^2}-1}}
&=\frac{2}{r}\sqrt{r^2/p_x^2-1}\Big|_{r_*}^{r_{\infty}}=\frac{2}{p_x}.
\ea
It is clear that
\be
L_{AdS}=2\log(2r_{\infty})+2\log\frac{l_{AdS}}{2}.
\ee

Similarly in BTZ, with $E=0$, and set $m=1$,
\be
\dot{r}^2= (r^2-1)(1-\frac{p_x^2}{r^2}), \hspace{1cm}
r'^2=(r^2-1)r^2\left(\frac{r^2}{p_x^2}-1\right), \hspace{1cm}
r_*=p_x.
\ee
\ba
L_{BTZ}&=&2\log(2r_{\infty})-2\log\sqrt{p_x^2-1}, \nonumber \\
l_{BTZ}&=&2~\mathrm{arctanh}(p_x)+i\pi.
\ea
One could find the relation,
\be
L_{BTZ}=2\log(2r_{\infty})+2\log\left(\sinh\left(\frac{l_{BTZ}}{2}\right)\right).
\ee
A crucial feature is $L_{BTZ}(l)>L_{AdS}(l)$ for any $l>0$.

If we shift a pure spacelike interval on the boundary, with fixed length $l$, along time direction from AdS region to BTZ region, when $t_b<0$, the corresponding geodesic is totally in AdS bulk. After it passes over $v=0$, the geodesic crosses the shell for twice. And if $t_b>\frac{l}{2}$, the geodesic will be totally in BTZ bulk. Since the ones totally in AdS or in BTZ bulk are pretty simple, we should focus on the ones crossing the shell below.

\begin{figure}[htbp]
\begin{center}
\includegraphics[scale=.8]{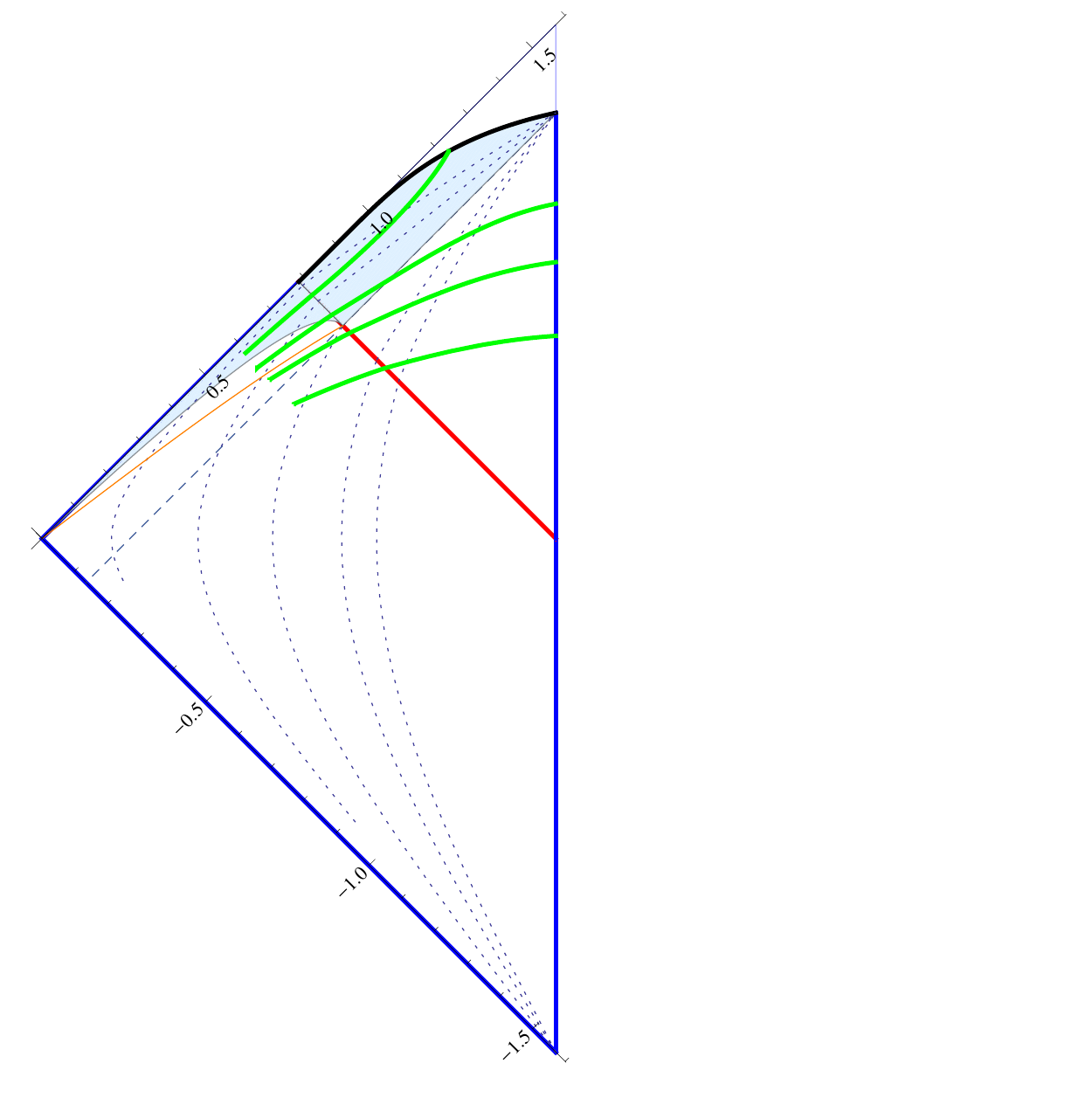}
\caption{Several geodesics(green curves) are shown in the figure. Any symmetric geodesic with the symmetric point in the shaded region falls to the singularity, while all the others will eventually approach the asymptotic AdS boundary. The constant time (orange) curve through the crossing point of the horizon and mass-shell is drawn for comparison. The second geodesic from top is interesting, that it passes through behind horizon region and reaches asymptotic AdS boundary. Some numeric detail is $v_0=-0.4$, and $p_x=\{0.5,0.65,0.75,0.95\}$ respectively for the geodesics from top to bottom.
}
\label{fig:Geodesics_Vaidya_ThinShell}
\end{center}
\end{figure}

In general, a symmetric spacelike geodesic crossing the shell is nontrivial. Its turning point is in AdS region, and after the shell the geodesic might probe region behind the horizon, and even falls to the singularity. In Fig.\ref{fig:Geodesics_Vaidya_ThinShell} several geodesics are plotted. Any symmetric spacelike geodesic with its turning point in the shaded region will falls to the singularity. Note after the infalling mass shell, the shaded region is defined by the horizon.

A geodesic with both endpoints anchored to a pure spacelike interval on the boundary has two free parameters, $(t_b=v_b,l)$. Suppose it crosses the shell $v=0$ at $r=r_c$, together with the conserved momentum $p_x=r^2\dot{x}$, the system can be described by the parameter combination $(p_x,r_c)$ as well.

Since the geodesic is symmetric, the part in AdS region has $E_A=r^2\dot{t}=0$, i.e. it is along a constant time curve. It is easy to find the position of the turning point, $(r_0,v_0)$,
\be
r_0=p_x, \hspace{1cm} v_0=t_0-\frac{1}{r_0}=\frac{1}{r_c}-\frac{1}{p_x}.
\ee
It is not hard to solve the geodesic equations,
\be
r=\half e^{-\tau}\left(e^{2\tau}+p_x^2\right).
\ee
It crosses the shell at $r=r_c$, with affine parameter value at the two crossing positions
\be
\tau^{(A)}_{c1,2}=\log(r_c\mp\sqrt{r_c^2-p_x^2}).
\ee
The ``\textit{A}'' used as superscript and subscript later indicates AdS, while ``\textit{B}'' for BTZ to clear up the confusion.
One can calculate the expansion along $x$ and the proper length,
\ba
\Delta x_A &=& \frac{-2p_x}{e^{2\tau}+p_x^2}\Big|_{\tau_{c1}}^{\tau_{c2}}=\frac{2}{r_c p_x}\sqrt{r_c^2-p_x^2}, \\
L_A &=& \tau^{(A)}_{c2}-\tau^{(A)}_{c1}=\log\left(\frac{r_c+\sqrt{r_c^2-p_x^2}}{{r_c-\sqrt{r_c^2-p_x^2}}}\right).
\ea
Since the geodesic is symmetric, one can focus on the half  part with $\tau$ increasing from $r=r_0$ to the boundary. Then at the interception with the shell,
\be
r_A'\Big|_{r_c}=\frac{r_c^2}{p_x}\sqrt{r_c^2-p_x^2}, \hspace{.5cm}
r_B'\Big|_{r_c}=\left(1-\frac{1}{2r_c^2}\right)r_A'\Big|_{r_c}, \hspace{.5cm}
E\equiv E_B=-\frac{1}{2r_c^2}\sqrt{r_c^2-p_x^2}.
\ee
$E\neq 0$ and the geodesic experiences a refraction.

Now let us turn to the BTZ part. With a nonzero $E$, the solution is
\be
\label{eqn:R1_BTZ}
R_1\equiv r^2=\frac{1}{4}\left(e^{\tau}+B_+e^{-\tau}\right)\left(e^{\tau}+B_-e^{-\tau}\right)
\equiv \frac{1}{4\alpha}\left(\alpha+B_+\right)\left(\alpha+B_-\right).
\ee
where $\alpha\equiv e^{2\tau}$, and
\be
\label{eqn:Bdefinition}
B_{\pm}=(p_x\pm 1)^2-E^2.
\ee
For simplicity one can introduce
\be
\label{eqn:Adefinition}
A_{\pm}=p_x^2-(1\pm E)^2,
\ee
and they satisfy
\be
A_+A_-=B_+B_-, \hspace{1cm} \frac{1}{4}(B_++B_-)-1=\frac{1}{4}(A_++A_-).
\ee
It is easy to verify
\be
r(\tau)^2-1 = \frac{1}{4}\left(e^{\tau}+A_+ e^{-\tau}\right)\left(e^{\tau}+A_- e^{-\tau}\right).
\ee
Some details of a general solution is discussed in section \ref{sec:NonPureSpacelikeAdSBTZ}.

At the crossing point,
\be
\alpha_c=e^{2\tau_c}=\half\left[-(B_++B_-)+4r_c^2\pm\sqrt{-4B_+B_-+(B_++B_--4r_c^2)^2}\right],
\ee
where $\pm$ is the sign of $r_B'\Big|_{r_c}$.
One can solve
\be
\Delta x_B=\half\log\left(\frac{B_++\alpha_c}{B_-+\alpha_c}\right), \hspace{.5cm}
\Delta t_B=\half\log\left(\frac{A_-+\alpha_c}{A_++\alpha_c}\right), \hspace{.5cm}
L_B=-\tau_c, \hspace{.5cm}
t_b=v_b=\Delta t_B+\textrm{arctanh}\frac{1}{r_c}.
\ee
Combining all the three parts together,
\ba
l_x &=& \Delta x_A+2\Delta x_B= \frac{2}{r_c p_x}\sqrt{r_c^2-p_x^2} + \half\log\left(\frac{B_++\alpha_c}{B_-+\alpha_c}\right), \\
L_{reg} &=& L_A+2L_B= \log\left(\frac{r_c+\sqrt{r_c^2-p_x^2}}{{r_c-\sqrt{r_c^2-p_x^2}}}\right)-\log\alpha_c.
\ea
We have already regularized the proper length by removing the common infinity part $2\log(2r_{\infty})$.
It might be useful to write the boundary time value as
\be
e^{2t_b}=\frac{(A_-+\alpha_c)(r_c+1)}{(A_++\alpha_c)(r_c-1)}.
\ee
To verify SSA of pure spacelike intervals on boundary, we need to keep $t_b$ fixed, and check the relation between $L_{reg}$ and $l_x$.

\begin{figure}[htbp]
\begin{center}
\includegraphics[scale=1.5]{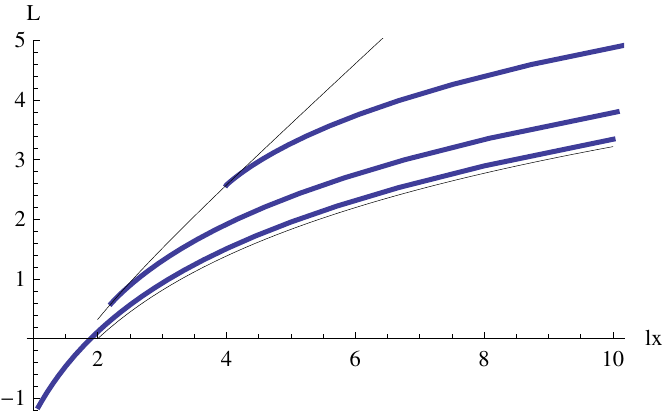}
\caption{$L(l_x)$ for pure spacelike interval at $t_b=\{0.5,1.1,2\}$ on the boundary, respectively from bottom to top. The black curves are $L(l_x)$ functions for BTZ and pure AdS, from top to bottom respectively.
}
\label{fig:PureSpacelikeIntervalsVaidyavB}
\end{center}
\end{figure}

Unfortunately it is hard to derive a clean formula of $L(l_x)$, so we turn to numeric solution. In Fig.\ref{fig:PureSpacelikeIntervalsVaidyavB}, we give a $L(l_x)$ function example with $t_b=0.5$ on the asymptotic AdS boundary. The proper length $L$ increases monotonically with $l_x$. When $l_x$ is small, it behaves like BTZ space, and the large $l_x$ behavior is determined by AdS part, which is reasonable, since small expansion geodesic is mainly in BTZ bulk while large expansion ones are mainly in AdS space.

To verify SSA inequalities, one can define
\ba
\label{eqn:SSAdefinitionI12}
I_1(A,B,C) &\equiv& S_{AB}+S_{BC}-S_A-S_C, \\
I_2(A,B,C) &\equiv& S_{AB}+S_{BC}-S_{ABC}-S_B.
\ea
If both $(I_1,I_2)$ are nonnegative, SSA is satisfied.

\begin{figure}[htbp]
\begin{center}
\includegraphics[scale=1.5]{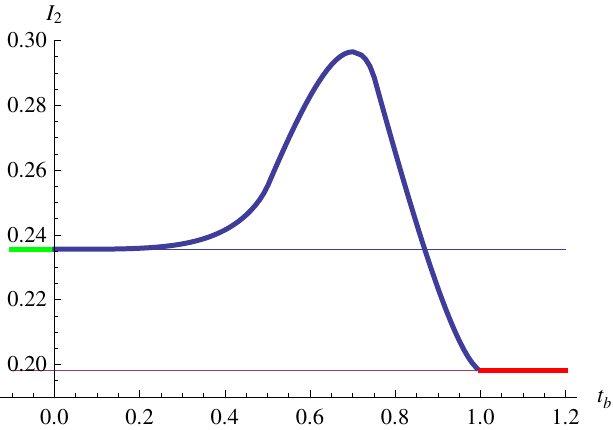}
\caption{With $(l_A,l_B,l_C)=(0.5,1,0.5)$, we move the pure spacelike intervals on boundary from AdS to BTZ. $I_2$ is always positive. The green part is for AdS, and the red part is for the case that all the relevant geodesics are in BTZ space.
}
\label{fig:PureSpacelikeIntervalsVaidyaI2vB}
\end{center}
\end{figure}

The $L(l_x)$ function in Fig.\ref{fig:PureSpacelikeIntervalsVaidyavB} tells us that it is also concave, and thus SSA is satisfied. In addition, one can check numerically of $I_2$ function, as in Fig.\ref{fig:PureSpacelikeIntervalsVaidyaI2vB}. We fix $(l_A,l_B,l_C)=(0.5,1,0.5)$ and move the pure spacelike intervals on boundary from AdS to BTZ. The bump behavior of $I_2$ indicates that SSA is always satisfied.\footnote{Thick-shell Vaidya is studied in \cite{AbajoArrastia:2010yt}, the monotonicity and concavity of $L(l)$ are apparent in Fig. 3 of it, as well as SSA. }

\subsection{Negative-energy Vaidya spacetime}

In physical Vaidya space, a spacelike geodesic could probe from BTZ region into AdS region, and experience a flatter function of $L(l_x)$. This induces us the idea that how the function $L(l_x)$ should behave if the two bulk regions are exchanged. It turns out the exchange is connected with the violation of NEC, and SSA is violated.

For a general Vaidya space,
\be
ds^2=-f(r,v)dt^2+\frac{dr^2}{f(r,v)}+r^2dx^2 =-f(r,v)dv^2+2drdv+r^2dx^2,
\ee
where
\be
f(r,v)=r^2-m(v).
\ee
The NEC requires
\be
T_{\mu\nu}n^{\mu}n^{\nu}\geq 0,
\ee
where $n^{\mu}$ is lightlike, i.e. $n^{\mu}n_{\mu}=0$.
In a Vaidya space with asymptotically $AdS_{d+1}$ boundary, the Einstein equation is
\be
G_{\mu\nu}+\Lambda g_{\mu\nu}=T_{\mu\nu}, ~\mathrm{where~}\Lambda=-\frac{d(d+1)}{2l^2},
\ee
here $l$ is the AdS radius, and $\Lambda=-3$ for $d=2$ if setting $l=1$.
Suppose a null vector is $n^{\mu}=(a,b,0)$, the $x$ component is set to zero for convenience. Solving the null normal equation $n^{\mu}n_{\mu}=0$ gives
\be
a=0,\hspace{1cm}\mathrm{or~~~}b=f(r,v)a/2.
\ee
The first one is simply just $n^{\mu}\sim (0,1,0)$ exactly along coordinate $r$, while the second $n^{\mu}\sim (1,f/2,0)$.
The first one is trivial since $G_{rr}=T_{rr}=0$, while the second has,
\be
T_{\mu\nu}n^{\mu}n^{\nu}=G_{\mu\nu}n^{\mu}n^{\nu}=\frac{m'}{2r}.
\ee
Thus the NEC is equivalent to $m'(v)\geq 0$.

If we turn on a negative-energy mass shell, with $m'(v)\leq 0$, it violates the NEC and  results into a negative-energy Vaidya spacetime: a BTZ evolves into a pure AdS spacetime.

For an observer sitting on the asymptotic boundary, if interested only in the pure spacelike intervals with expansion along $x$ of $l_x$, there are basically three kinds of spacelike geodesics.
As usual, suppose the thin shell is emitted from the boundary at $v=0$, and the observer is at $t=t_b$.
\begin{enumerate}
\item
If $t_b<0$, the geodesics is totally in BTZ region.
\item
For $t_b>0$, if  $l_x<\bar{l}_x(t_b)$, it is totally in $AdS_3$.
\item
For $t_b>0$, if  $l_x>\bar{l}_x(t_b)$, the geodesic crosses the mass shell, and approaches BTZ region.
\end{enumerate}
It is easy to find $\bar{l}_x(t_b)=2t_b$, which is the same as in physical Vaidya space.

\begin{figure}[htbp]
\begin{center}
\includegraphics[scale=.8]{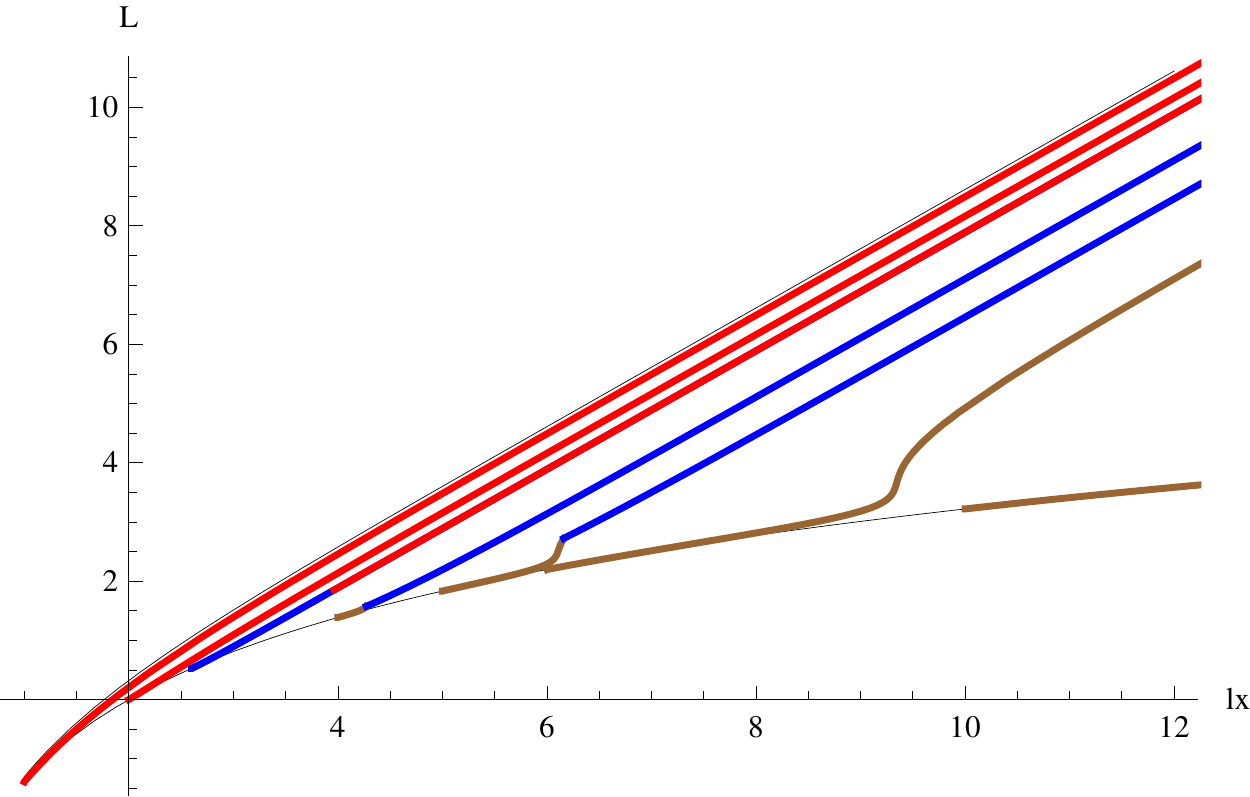}
\caption{A series of geodesics ending on AdS boundary at $t_b=\{0.5, 1, 1.3, 2, 2.5, 3, 5\}$ from top to bottom respectively. Red curves are calculated from $r_c>1$, the blue ones are from $\half<r_c^2<1$, and the brown come from $r_c^2<\half$. Clearly the middle three with $t_b=\{1.3,2,2.5\}$ are combination from two cases. The top and bottom black curves are given for BTZ and pure $AdS_3$ respectively.
}
\label{fig:UnphysicalVaidyaLoflxlist}
\end{center}
\end{figure}

We shall keep the details in appendix \ref{appendix:unphysicalVaidya}, and list some features here.
The calculation is complicated, and the geodesic takes different branches of the solution when the pure spacelike interval on boundary changes with parameters $(t_b,l_x)$.
In Fig.\ \ref{fig:UnphysicalVaidyaLoflxlist}, we plot a series of geodesics with different boundary time value of $t_b$. A general behavior is when $t_b$ increases, the curve moves from BTZ to AdS. From the value ranges of $t_b$, (\ref{eqn:vBconstraintrc>1},\ref{eqn:vBconstraintrc2_Half_1},\ref{eqn:vBconstraintrc2_less_Half}), there are some overlaps. The combinations give continuous $L(l_x)$ curves. An important property is although the curves are still monotonic, the concavity is not preserved, and this results into the violation of strong subadditivity.

Keeping $t_b>0$ fixed, the spacelike geodesics with small $l_x$ feels only the effect of $AdS_3$ bulk, while the ones with large $l_x$ cross the shell, and eventually the BTZ's effect dominates if $l_x$ is large enough.

In both Fig.\ \ref{fig:UnphysicalVaidyaLoflxlist} and Fig.\ \ref{fig:PureSpacelikeIntervalsVaidyavB}, for a pure spacelike interval on the boundary, the small $l_x$ behavior is determined by the bulk after the mass shell, while the large $l_x$ is dominated by the one before the shell. Since $L_{BTZ}(l)>L_{AdS}(l)$, only the $L(l)$ curve for the negative-energy Vaidya contains a non-concave part, and thus SSA is violated. Heuristically we could make a conjecture. Suppose there are two adjacent regions $A$ and $B$, sharing the same asymptotic boundary. For spacelike geodesic with its endpoints attached on the asymptotic boundary, $L_B(l)>L_A(l)$. Then for any such spacelike geodesic probing from bulk region $A$ into $B$, SSA is violated. On the other hand, if the geodesic is probing from $B$ into $A$, SSA is satisfied. A manifest possible example is a domain wall at constant radius, the two sides of which might have very different behavior, say one is AdS while the other is BTZ, or one is AdS and the other is Lifshitz-like space, or the two have different cosmological constants. Naively to think, the violation of SSA is determined by properties of the two regions, while the domain wall might impose some effect, e.g.\ a negative tension might change the story from with a positive tension. (This may be related to the fact that proposed entanglement-entropy-based holographic $c$-theorems make use of the NEC, albeit in a static context \cite{Liu:2012ee,Myers:2012ed}.) However, to make the system stable, NEC might have to be satisfied.

\section{General spacelike intervals}

In this section we expand our considerations to include spacelike intervals that do not lie in a constant-time surface (with respect to the Lorentz frame distinguished by the system's spatial translation symmetry), a case that has been somewhat neglected in the literature so far. In particular, in order to check SSA thoroughly we consider a variety of possible configurations, in which the intervals are not only not constant-time but not collinear with each other (see Fig.\ \ref{fig:3spatialCases}).

It is instructive first to consider the case of general spacelike intervals in a CFT at finite temperature. Since the state is translationally invariant in both space and time, the entropy is a function of $\Delta x$ and $\Delta t$.\footnote{From now on we will use $\Delta x$ instead of $l_x$ as in pure spacelike case in previous section, to avoid the potential confusion to the length $l=\sqrt{(\Delta x)^2+(\Delta t)^2}$ on the boundary.} Defining null coordinates $u=x-t$, $v=x+t$, the entropy can be expressed in terms of the function $s$ defined in \eqref{nonzeroT} \cite{Hubeny:2007xt}
\begin{equation}\label{nonzeroTgeneral}
S = \frac12\left(s(\Delta u) + s(\Delta v)\right).
\end{equation}
(The same formula applies at zero temperature, with $s$ given by \eqref{zeroT}.) Equation \eqref{nonzeroTgeneral} clearly obeys SSA, since each term $s(\Delta u)$ and $s(\Delta v)$ obeys it separately. It was shown in \cite{Hubeny:2007xt} that the HRT formula, applied to the BTZ spacetime, reproduces \eqref{nonzeroTgeneral}.

Note that equation \eqref{nonzeroT} has the following interesting feature (for any concave function $s$). If we consider the quantity $I_2$ defined in \eqref{eqn:SSAdefinitionI12}, which measures how far \eqref{SSA1} is from being saturated, it can be written
\begin{equation}
I_2 = \frac12(I_2^u+I_2^v)\,,
\end{equation}
where
\begin{equation}
I_2^u = s(\Delta u_A+\Delta u_B)+s(\Delta u_B+\Delta u_C) - s(\Delta u_A+\Delta u_B+\Delta u_C) - s(\Delta u_B)
\end{equation}
and similarly for $I_2^v$. Now, in the limit that $A$ is null, either $\Delta u_A\to0$ or $\Delta v_A\to0$ (depending on which way it is tilted), implying either $I_2^u\to0$ or $I_2^v\to0$. If we also take the limit where $C$ is null, but tilted the other way---as in the trapezoid configuration of Fig.\ \ref{fig:3spatialCases}---then \emph{both} $I_2^u$ and $I_2^v$ will go to 0, so $I_2\to0$ and SSA will be saturated. (Note that, even when both $A$ and $C$ are null, all of the intervals appearing in the definition of $I_2$ are spacelike.) In the rest of this section we will consider a Vaidya spacetime that makes a transition from AdS to BTZ (or vice versa for negative-energy Vaidya). We will pay particular attention to the trapezoid configuration with $A$ and $C$ close to null: since SSA is just barely satisfied in both AdS and BTZ for this configuration, it offers the best chance of finding a violation if one exists.

\begin{figure}[htbp]
\begin{center}
\includegraphics[scale=.5]{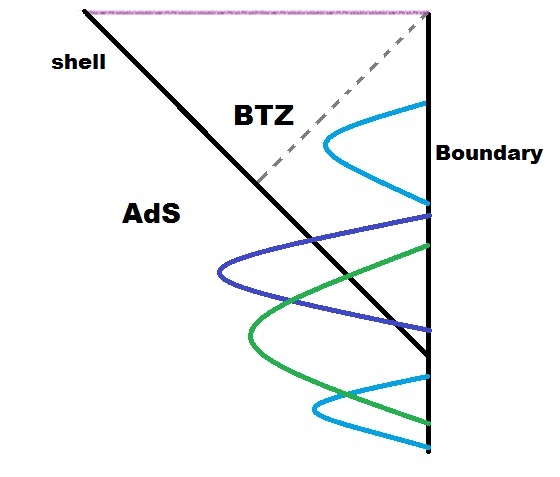}
\caption{The four basic kinds of geodesic with both endpoints attached on the asymptotic AdS boundary. The geodesic might be totally in AdS bulk or BTZ space, or it might crosses the thin-shell once or twice. Note that the shell-crossing geodesics may reach the region behind the horizon (not shown in the figure).
}
\label{fig:Vaidya_TheCrossings}
\end{center}
\end{figure}

As shown in Fig. \ref{fig:Vaidya_TheCrossings}, for a spacelike geodesic with both endpoints attached on the boundary, there are four possible configurations: it could be totally in AdS or BTZ, or it might cross the thin-shell once or twice. We will discuss these possibilities in subsections 4.1 and 4.2, finding exact solutions in each case. Then in subsection 4.3 we will put together the solutions in order to test SSA. This involves a certain amount of numerical work, since the formulae for the proper lengths of the geodesics in terms of boundary data are complicated and implicit. We find that the infalling matter actually has the effect of increasing $I_2$ for the trapezoidal configuration, as long as it has positive mass (see Fig.\ \ref{fig:VaidyaSSATrapzoidI}), so SSA is safe; nor do we find violations for $I_1$, or for either $I_1$ or $I_2$ for any other configurations. On the other hand, we do find that negative-energy infalling matter does drive $I_2$ negative (see Fig.\ \ref{fig:UnphyVaidyaSSATZI}), showing that violation of SSA is correlated with violation of the null energy condition (NEC).

\subsection{Geodesics totally in pure AdS or BTZ}
\label{sec:NonPureSpacelikeAdSBTZ}

The formula in pure AdS or BTZ space is simple. One can easily find, in pure AdS, the solution is
\be
r=\half e^{-\tau}\left(e^{2\tau}+p_x^2-E^2\right),
\ee
where the conserved momenta are defined
\be
E=r^2\dot{t}, \hspace{1cm} p_x=r^2\dot{x}.
\ee
The boundary has the affine parameter values
\be
\tau_{\infty}=\ln{2r_{\infty}}, \hspace{1cm}
\tau_{-\infty}=-\ln{2r_{\infty}}+\ln(p_x^2-E^2),
\ee
while the turning point is at
\be
r_{turning}^2=p_x^2-E^2<p_x^2
\ee
instead of $r=p_x$ as with $E=0$ case, and
\be
\tau_{turning}=\half\ln(p_x^2-E^2).
\ee
One could find the time values with $(p_x,E,v_0)$, where $v_0$ is the $v$ value of the turning point,
\be
t_0=v_0+\frac{1}{\sqrt{p_x^2-E^2}},~~
t_{max}=t_0+\frac{E}{p_x^2-E_0^2}, ~~
t_{min}=t_0-\frac{E}{p_x^2-E_0^2}.
\ee
and the expansion on boundary,
\ba
\Delta x= \frac{2p_x}{p_x^2-E^2},  \hspace{1cm} &\Delta t=\frac{2E}{p_x^2-E^2}, \\
L=2\ln{2r_{\infty}}-\ln(p_x^2-E^2) ~~&\sim \ln(\Delta x+\Delta t)+\ln(\Delta x-\Delta t).
\label{eqn:Lofxt_pureAdS}
\ea

The discussion in BTZ space is slightly complicated. For the geodesic equations
\ba
\dot{r}^2= f+E^2-\frac{f}{r^2}p_x^2, ~~r'^2=\frac{r^4f}{p_x^2}+\frac{r^4}{p_x^2}E^2-r^2f,
~~E=f\dot{t},~~ p_x=r^2\dot{x},
\ea
there are two branches of solution,
\ba
\label{eqn:R1_BTZ_generalE}
R_1(\tau)=r^2(\tau)&=&\frac{1}{4}\left(e^{\tau}+B_+e^{-\tau}\right)\left(e^{\tau}+B_-e^{-\tau}\right), \\
R_2(\lambda)=r(\lambda)^2 &=& -\frac{1}{4}\left(e^{\lambda}B_+-e^{-\lambda}\right)\left(e^{\lambda}B_--e^{-\lambda}\right),
\label{eqn:R2_BTZ_generalE}
\ea
here we used the function $B_{\pm}$ introduced in equation (\ref{eqn:Bdefinition}), and list them here to be clear,
\ba
B_{\pm}=(p_x\pm 1)^2-E^2, \hspace{1cm} A_{\pm}=p_x^2-(1\pm E)^2.
\ea
The solution $R_1(\tau)$ was derived in \cite{Balasubramanian:2011ur}.
If we introduce $\alpha\equiv e^{2\tau}$ or $\alpha\equiv e^{2\lambda}$, the solutions are simplified into
\be
R_1(\alpha)=\frac{1}{4\alpha}(B_++\alpha)(B_-+\alpha), \hspace{1cm}
R_2(\alpha)=-\frac{1}{4\alpha}(\alpha B_+-1)(\alpha B_--1).
\ee
With the quantities $A_{\pm}$, one has
\ba
R_1(\tau)-1 &=& \frac{1}{4}\left(e^{\tau}+A_+ e^{-\tau}\right)\left(e^{\tau}+A_- e^{-\tau}\right), \\
R_2(\lambda)-1 &=& -\frac{1}{4}\left(e^{\lambda}A_+-e^{-\lambda}\right)\left(e^{\lambda}A_--e^{-\lambda}\right).
\ea
Thus the signs of $B_{\pm}$ tell us whether the geodesic falls to the singularity $r=0$, while the signs of $A_{\pm}$ indicate whether it crosses the horizon $r=1$.

\begin{figure}[htp]
\subfigure[$p_x=1.5,E=1$]{
\includegraphics[scale=0.27]{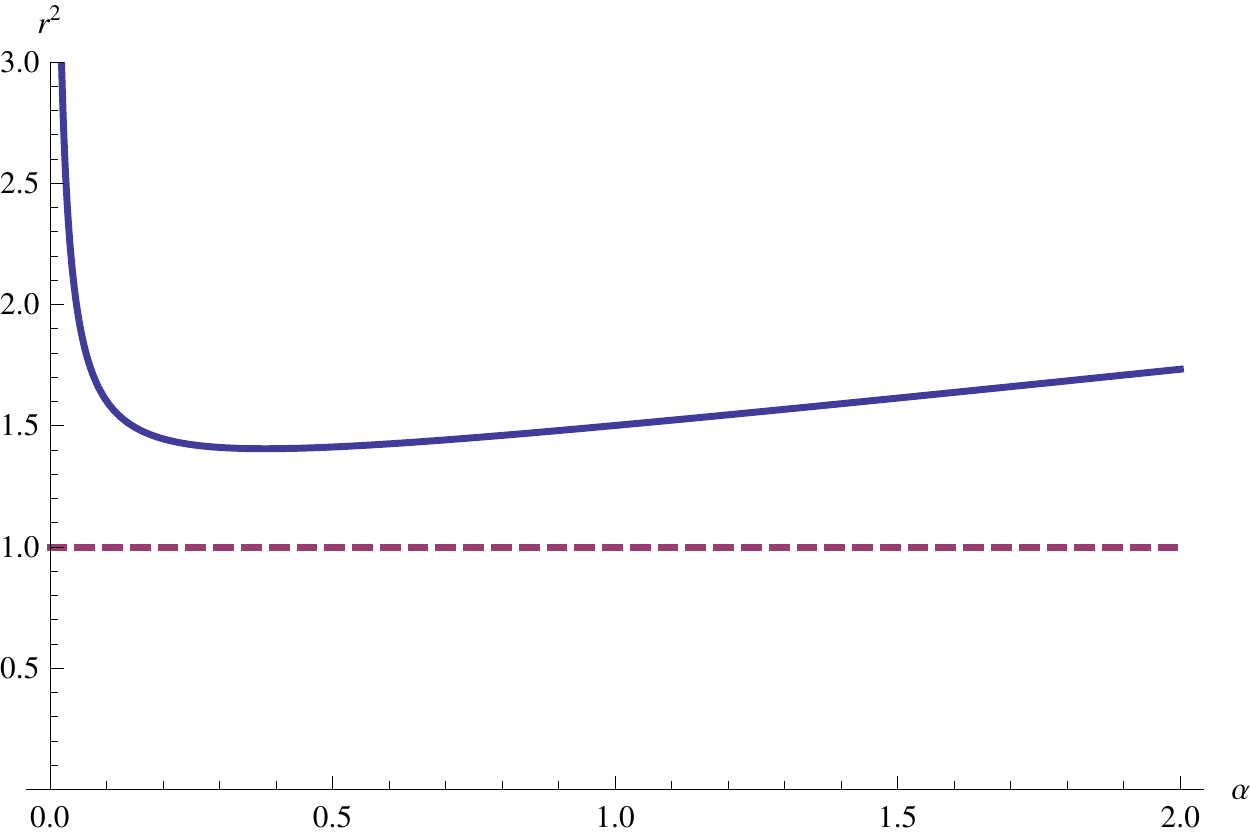}
\label{fig:R1_alpha_cases_outsidehorizon}
}
\subfigure[$p_x=1.2,E=0.1$]{
\includegraphics[scale=0.27]{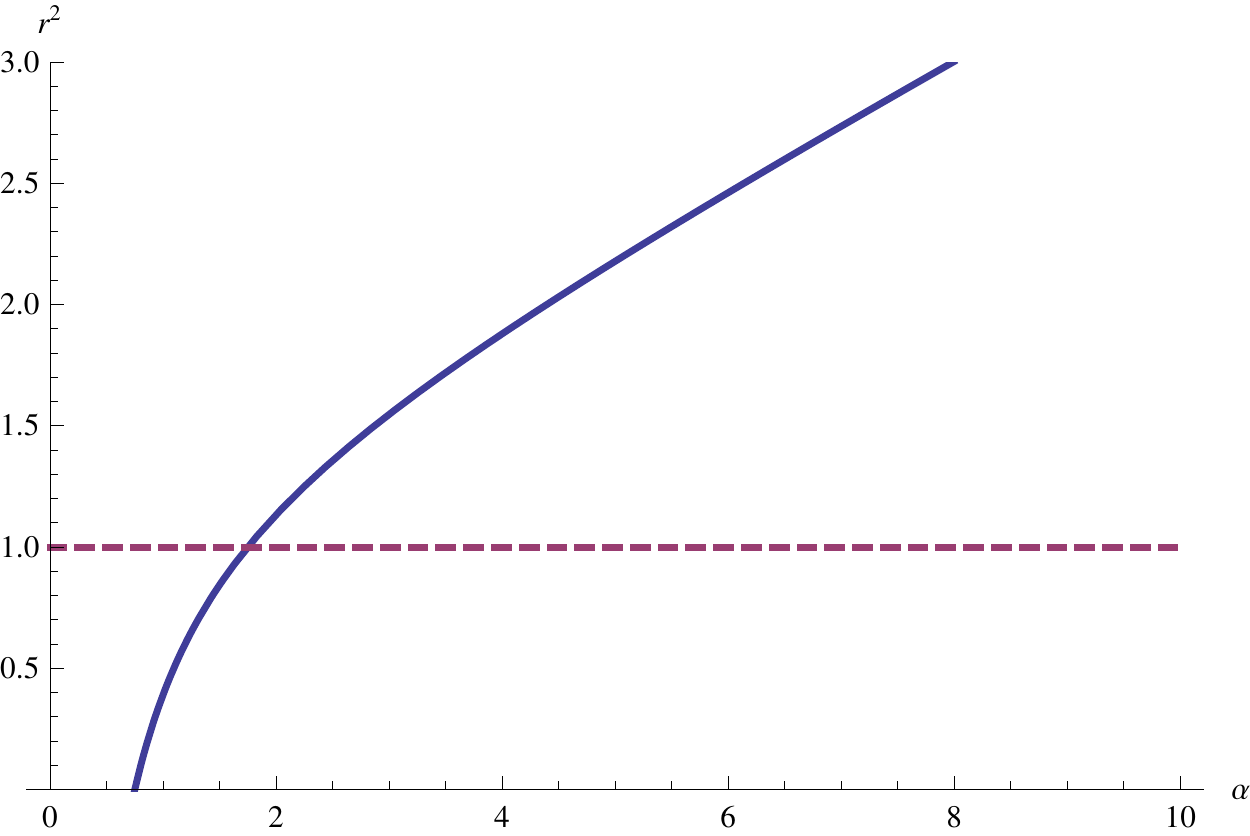}
\label{fig:R1_alpha_cases_mono}}
\subfigure[$p_x=0.1,E=0.5$]{
\includegraphics[scale=0.27]{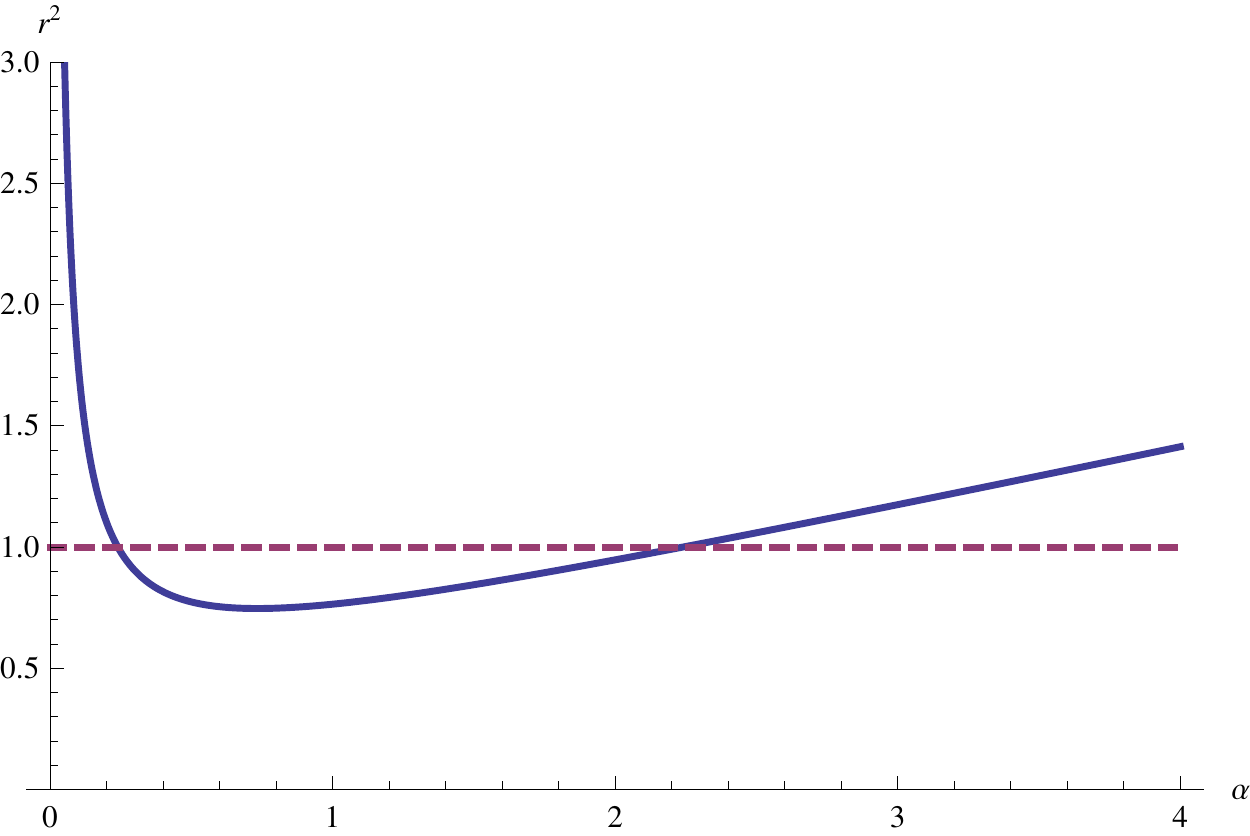}
\label{fig:R1_alpha_cases_2horizon}}
\subfigure[$p_x=0.5,E=2$]{
\includegraphics[scale=0.27]{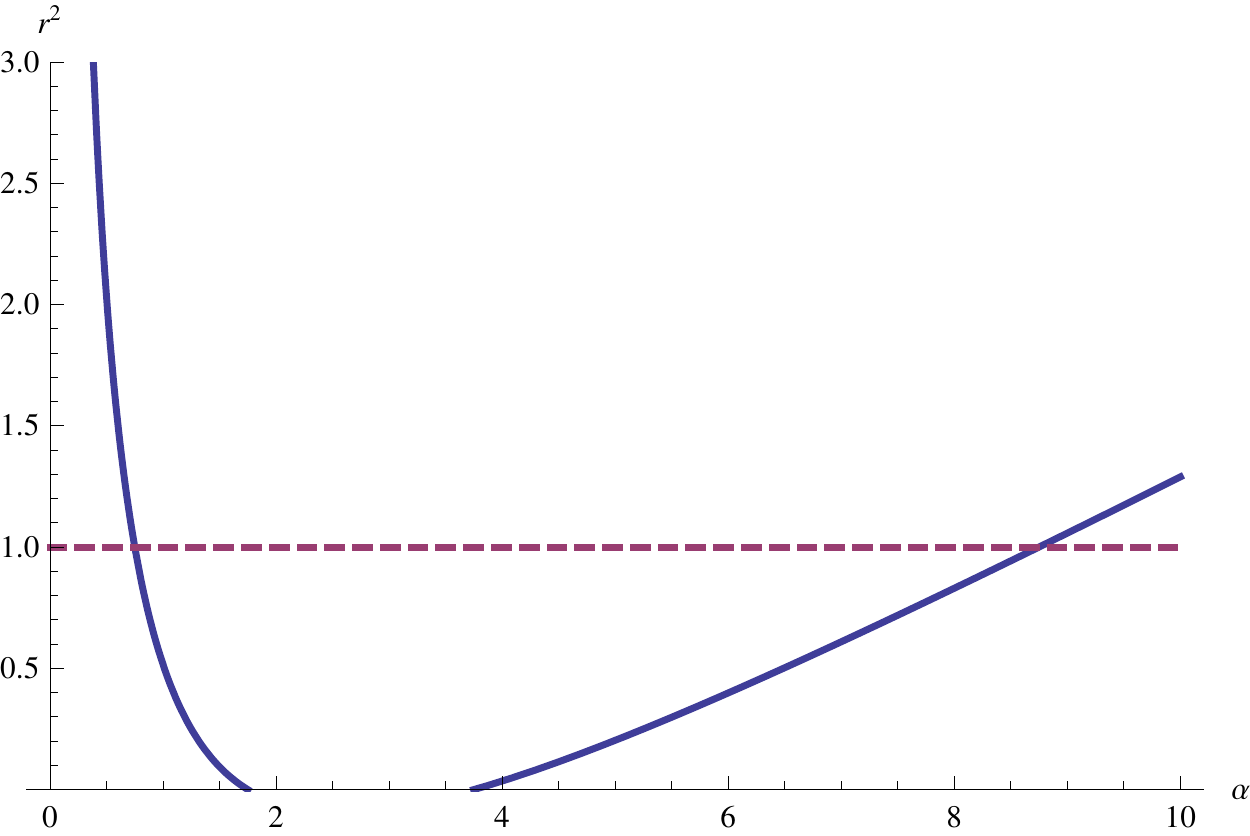}
\label{fig:R1_alpha_cases_2horizon_2sing}}
\caption{The four cases of the solution $R_1(\alpha)$. The dashed line is the horizon.}
\label{fig:R1_alpha_cases}
\end{figure}

An analysis of the solutions $R_{1,2}$ by quantities $A_{\pm},B_{\pm}$ is illustrative. The solution $R_1(\tau)$ has at least one limit $\tau\ra\pm\infty$ attached to asymptotic AdS boundary, and can be classified into four different kinds(Fig.\ref{fig:R1_alpha_cases}):
\begin{enumerate}
\item
If $A_{\pm},B_{\pm}>0$, the geodesic is totally outside of the horizon(Fig.\ref{fig:R1_alpha_cases_outsidehorizon}).
\item
If $A_+A_-=B_+B_-<0$, the geodesic crosses the horizon and reaches the singularity(Fig.\ref{fig:R1_alpha_cases_mono}).
\item
If $A_{\pm}<0,B_{\pm}>0$, the geodesic crosses the horizon twice, and goes back to AdS boundary(Fig.\ref{fig:R1_alpha_cases_2horizon}).
\item
If$A_{\pm},B_{\pm}<0$, the geodesic crosses the horizon twice and hits the singularity twice(\ref{fig:R1_alpha_cases_2horizon_2sing}).
\end{enumerate}

\begin{figure}[htp]
\centering
\subfigure[$p_x=1.1,E=0.2$]{
\includegraphics[scale=0.35]{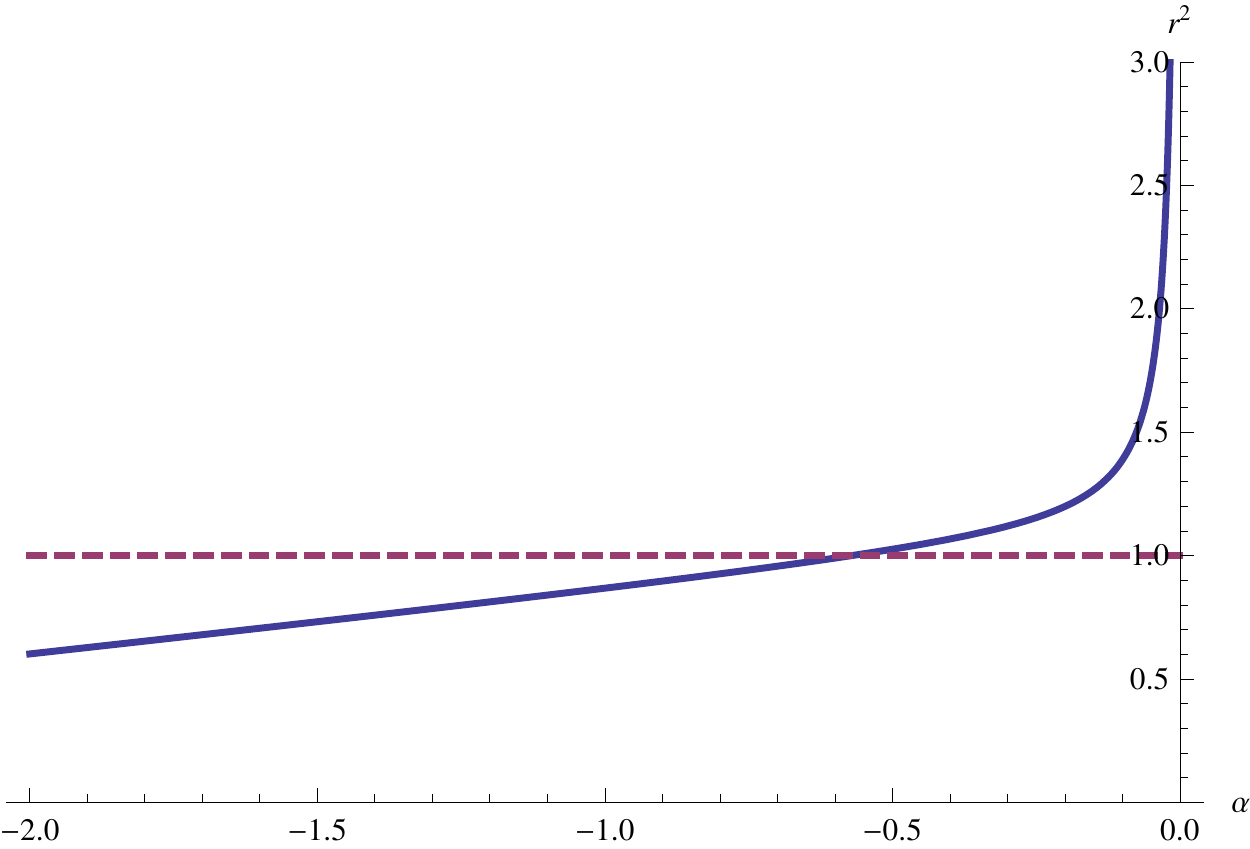}
\label{fig:R2_alpha_cases_mono}
}
\subfigure[$p_x=0.7,E=0.2$]{
\includegraphics[scale=0.35]{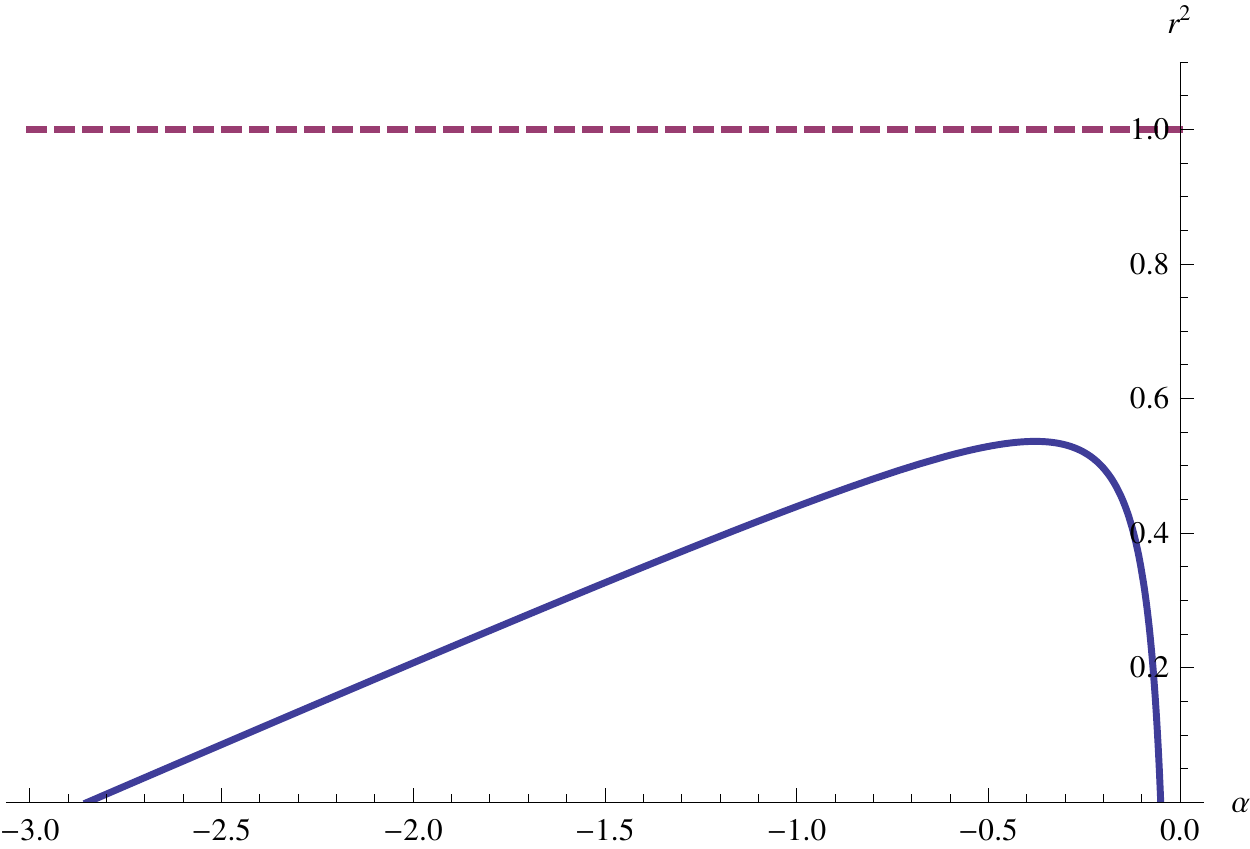}
\label{fig:R2_alpha_cases_insidehorizon}}
\subfigure[$p_x=1.9,E=0.8$]{
\includegraphics[scale=0.35]{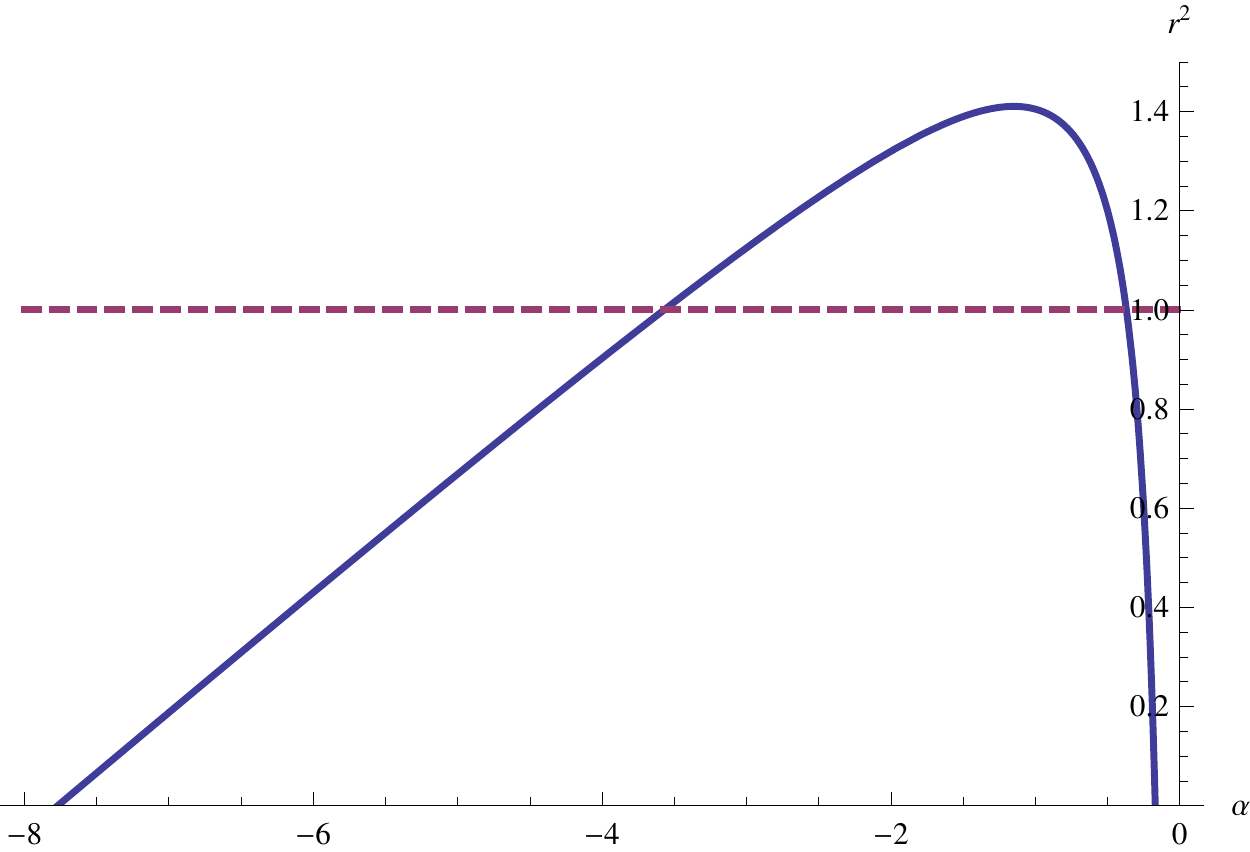}
\label{fig:R2_alpha_cases_2horizon}}
\caption{The three physical cases of the solution $R_2(\alpha)$. The dashed line is the horizon.}
\label{fig:R2_alpha_cases}
\end{figure}

On the other hand, the solution $R_2(\alpha)$ describes the geodesics falling to the singularity(Fig.\ref{fig:R2_alpha_cases}):
\begin{enumerate}
\item
$A_+A_-=B_+B_-<0$, the geodesic crosses the horizon once, and the two endpoints approach the AdS boundary and the singularity respectively(Fig.\ref{fig:R2_alpha_cases_mono}).
\item
$B_{\pm}>0,A_{\pm}<0$, the geodesic is totally inside the horizon and both endpoints fall to the singularity(Fig.\ref{fig:R2_alpha_cases_insidehorizon}).
\item
$B_{\pm}>0,A_{\pm}>0$, part of the geodesic is outside of the horizon, while both endpoints fall to the singularity(Fig.\ref{fig:R2_alpha_cases_2horizon}).
\item
$B_{\pm}<0,A_{\pm}<0$, it gives that the value of $R_2$ is always negative, which is non-physical, and not shown in Fig.\ref{fig:R2_alpha_cases}.
\end{enumerate}

For a spacelike geodesic totally in BTZ space with both endpoints on boundary, it can take only the branch of solution $R_1$, with $B_{\pm}>0$, which is obvious from Fig.\ref{fig:R1_alpha_cases}.

Again it might be useful to write down the affine parameter values,
\be
\tau_{turning}=\frac{1}{4}\ln\left(B_+B_-\right), \hspace{.5cm}
\tau_{\infty}=\ln\left(2r_{\infty}\right),\hspace{.5cm}
\tau_{-\infty}=-\ln\left(2r_{\infty}\right)+\half\ln\left(B_+B_-\right).
\ee
In addition, the turning point is at
\ba
r_0= \half\left(\sqrt{B_+}+\sqrt{B_-}\right), \hspace{1cm} t_0=v_0+\mathrm{arctanh}\left(\frac{1}{r_0}\right),
\ea
 The interval on boundary has
\ba
t_{max}&=&t_0-\half\ln\left(\frac{A_++\sqrt{B_+B_-}}{A_-+\sqrt{B_+B_-}}\right),  \\
t_{min}&=&t_0-\left(\half\ln\left(\frac{A_++\sqrt{B_+B_-}}{A_-+\sqrt{B_+B_-}}\right)-\half\ln\left(\frac{A_+}{A_-}\right)\right).
\ea
And the expansion is
\ba
\Delta x=\half\ln\frac{B_+}{B_-}, \hspace{1cm}& 
\Delta t=\half\ln\frac{A_-}{A_+}, \\
L =2\ln(2r_{\infty})-\half\ln\left(B_+B_-\right) ~~&  \sim \ln\left(\sinh\left(\frac{\Delta x+\Delta t}{2}\right)\right)+\ln\left(\sinh\left(\frac{\Delta x-\Delta t}{2}\right)\right)
\label{eqn:Lofxt_BTZ}
\ea

\subsection{Geodesics crossing the shell}

For a general spacelike geodesic crossing the shell, the solution has to satisfy the equations of motion (\ref{eqn:EOM_Vaidya_randv}).
Since the geodesic might reach the region behind BTZ's horizon, it is hard to tell which branch of solution, $R_1$ or $R_2$ it takes at a given radius $r$. One should be careful to check the evolution of geodesic with the affine parameter. In the following, we argue that for the geodesics we are interested, branch $R_1$ is complete to describe them. With the affine parameter transformation, the only section of $R_2$ connecting to asymptotic boundary is 1-1 mapping to the section of $R_1$ with $B_+B_-<0$. In addition, the discussion on 2nd derivative below justifies the $R_1$ solution.

Define $\alpha\equiv e^{2\tau}$, one finds that
\[
\left\{\begin{array}{ll}
R_1(\alpha)=\frac{1}{4\alpha}\left(\alpha+B_+\right)\left(\alpha+B_-\right) \\
R_2(\alpha)=-\frac{1}{4\alpha}\left(\alpha B_+-1\right)\left(\alpha B_--1\right)
\end{array}\right.
~~\Longrightarrow ~~
\left\{\begin{array}{ll}
R_1''(\alpha)=\half\frac{B_+B_-}{\alpha^3} \\
R_2''(\alpha)=-\half\frac{1}{\alpha^3}<0
\end{array}\right.
\]
For solution $R_1$, only the section with $A_+A_-=B_+B_-<0$ gives negative $R_1''$, while all sections of $R_2$ have $R_2''<0$.
Since we are only interested in the geodesics with both endpoints on AdS boundary, the part in AdS must have $r''(\tau)>0$, i.e. if introducing $\beta=e^{\tau}$,
\be
r^{(A)}=\frac{1}{2\beta}\left(\beta^2+p_x^2-E_A^2\right), ~~\Longrightarrow ~~
\frac{d^2r}{d\beta^2}=\frac{p_x^2-E_A^2}{\beta^3}>0 ~~\Longrightarrow ~~ p_x^2>E_A^2.
\ee
On the thin-shell,
\[
r_B''=\left(1-\frac{1}{2r_c^2}\right)r_A''  ~~\Longrightarrow ~~
\left\{\begin{array}{ll}
\mathrm{If~}r_c^2>\half, r_B''>0 ~\Longrightarrow~ R_1 \\
\mathrm{If~}r_c^2<\half, r_B''<0, \mathrm{and~}r_A'>0, r_B'<0.
\end{array}\right.
\]
Thus we say $R_1$ is enough to describe the part of geodesic in BTZ region.

One can derive the following matching equations,
\ba
v'=2(r_A'-r_B'),
\ea
and in AdS,
\be
v'=t'+\frac{r_A'}{r^2}=\frac{E_A}{p_x}+\frac{r_A'}{r^2} ~~\Longrightarrow
 r_B'=\left(1-\frac{1}{2r_c^2}\right)r_A'-\frac{E_A}{2p_x},
\ee
while in BTZ,
\be
v'=t'+\frac{r_B'}{r^2-1}=\frac{r^2}{r^2-1}\frac{E_B}{p_x}+\frac{r_B'}{r^2-1},
\ee
and this gives
\be
E_B=E_A-\frac{E_A}{2r_c^2}-\frac{p_x}{2r_c^4}r_A',
\ee
i.e. $E_B<E_A$.

\subsubsection{Crossing the shell once}

For a geodesic with both endpoints attached to AdS boundary, and crossing the thin-shell once, the general behavior is that it starts from its endpoint, falls into AdS bulk, turns back at its turning point, then crosses the thin-shell. After the thin-shell, if $r_B'|_{r_c}>0$, the geodesic goes straightly to the AdS boundary, otherwise it falls back to the singularity but turns back at some place, then goes to the AdS boundary. The sign of $r_B'|_{r_c}$ is determined by $(p_x,E_A,r_c)$. We list below some details of the simplest case, with $r_B'|_{r_c}>0$. The part in AdS has,
\be
\tau_c^{(A)}=\ln\left(r_c+\sqrt{r_c^2+E_A^2-p_x^2}\right), \hspace{1cm}
\tau_{-\infty}^{(A)}=\ln\left(r_{\infty}-\sqrt{r_{\infty}^2+E_A^2-p_x^2}\right),
\ee
and
\ba
l_x^{(A)} &=& \frac{-p_x}{r_c\left(r_c+\sqrt{r_c^2+E_A^2-p_x^2}\right)}+\frac{2p_x}{p_x^2-E_A^2},  \\
t^{(A)} &=& \frac{-E_A}{r_c\left(r_c+\sqrt{r_c^2+E_A^2-p_x^2}\right)}+\frac{2E_A}{p_x^2-E_A^2},  \\
L^{(A)} &=&\tau_c^{(A)}-\tau_{-\infty}^{(A)}.
\ea
The part in BTZ has
\be
\alpha_c^{(B)}=e^{2\tau_c^{(B)}}=\half\left[-(B_++B_-)+4r_c^2\pm\sqrt{-4B_+B_-+(B_++B_--4r_c^2)^2}\right],
\ee
and
\be
\tau_{\infty}^{(B)}=\ln(2r_{\infty}).
\ee
Some of the other quantities are
\ba
l_x^{(B)}=\half\ln\frac{B_++e^{2\tau_c^{(B)}}}{B_-+e^{2\tau_c^{(B)}}},  \hspace{0.5cm}
t^{(B)}=\half\ln\frac{A_-+e^{2\tau_c^{(B)}}}{A_++e^{2\tau_c^{(B)}}},  \hspace{0.5cm}
L^{(B)} =\tau_{\infty}^{(B)} -\tau_c^{(B)}.
\ea
The quantities $A_{\pm},B_{\pm}$ are defined by $(p_x,E_B)$.

The total expansion on boundary has
\be
\Delta x=l_x^{(A)}+l_x^{(B)}, \hspace{1cm}
\Delta t=t^{(A)}+t^{(B)},
\ee
and the total proper length
\be
L=L^{(A)}+L^{(B)}.
\ee

\subsubsection{Crossing the shell twice}

Suppose the geodesic crosses the shell at $r=r_{c1}$ and $r=r_{c2}$ for the bottom and top half respectively,
it is easy to check that $r_{c1}>r_{c2}$, and one should calculate the three parts separately. Again it has three free parameters, say $(p_x,E_A,v_0)$.

For the central AdS part, the turning point is at $(r_0,v_0)$, where
\be
r_0=\sqrt{p_x^2-E_A^2}, \hspace{1cm} t_0=v_0+\frac{1}{r_0}.
\ee
The crossing radius $(r_{c1},r_{c2})$ satisfy
\ba
\frac{1}{r_{c1}}&=&t_0+E_A\left(\frac{1}{p_x^2-E_A^2}-\frac{1}{r_{c1}\left(r_{c1}-\sqrt{r_{c1}^2+E_A^2-p_x^2}\right)}\right),\\
\frac{1}{r_{c2}}&=&t_0+E_A\left(\frac{1}{p_x^2-E_A^2}-\frac{1}{r_{c2}\left(r_{c2}+\sqrt{r_{c2}^2+E_A^2-p_x^2}\right)}\right).
\ea
The change of $t$ and $x$ are
\be
\Delta t_A=\frac{1}{r_{c2}}-\frac{1}{r_{c1}}, \hspace{1cm} \Delta x_A=\Delta t_A*\frac{p_x}{E_A},
\ee
and the proper length
\be
L_A=\ln\left[r_{c2}+\sqrt{r_{c2}^2+E_A^2-p_x^2}\right]-\ln\left[r_{c1}-\sqrt{r_{c1}^2+E_A^2-p_x^2}\right].
\ee

At the crossing $r=r_{c1}$, $\dot{r}<0$, thus
\be
r_A'=-\frac{r_{c1}^2}{p_x}\sqrt{r_{c1}^2+E_A^2-p_x^2}, \hspace{1cm}
E_{B1}=E_A\left(1-\frac{1}{2r_{c1}^2}\right)-\frac{p_x}{2r_{c1}^4}r_A'.
\ee
To write the solution of part BTZ1, one needs to check the values of $B^{(1)}_{\pm}\equiv B_{\pm}(p_x,E_{B1})$.
\ba
\textrm{If~}B^{(1)}_{+}B^{(1)}_-<0,& ~r^2(\tau)=\frac{1}{4}(e^{-\tau}+B^{(1)}_+e^{\tau})(e^{-\tau}+B^{(1)}_-e^{\tau}), \\
 & ~\mathrm{with~}\tau_{-\infty}=-\ln(2r_{\infty}),~\tau_{\infty}=\ln(2r_{\infty})-\half\ln\left(B_+^{(1)}B_-^{(1)}\right) \nn\\
\textrm{If~}B^{(1)}_{+}B^{(1)}_->0,& ~r^2(\tau)=\frac{1}{4}(e^{\tau}+B^{(1)}_+e^{-\tau})(e^{\tau}+B^{(1)}_-e^{-\tau}), \\
 & ~\mathrm{with~}\tau_{-\infty}=-\ln(2r_{\infty})+\half\ln\left(B_+^{(1)}B_-^{(1)}\right),~\tau_{\infty}=\ln(2r_{\infty}). \nn
\ea
More specifically, if $B^{(1)}_{+}B^{(1)}_-<0$, one can introduce
\be
\alpha_{c1}\equiv e^{2\tau^{B}_{c1}}
 =\frac{1}{2B^{(1)}_{+}B^{(1)}_-}\left[-\left(B^{(1)}_+ +B^{(1)}_-\right)+4r_{c1}^2
  -\sqrt{-4B^{(1)}_{+}B^{(1)}_-+\left(B^{(1)}_+ +B^{(1)}_- -4r_{c1}^2\right)^2}\right],
\ee
and
\ba
\Delta x_{B1}=\half\ln\left[\frac{1+B^{(1)}_+\alpha_{c1}}{1+B^{(1)}_-\alpha_{c1}}\right], \hspace{.5cm}
\Delta t_{B1}=\half\ln\left[\frac{1+A^{(1)}_-\alpha_{c1}}{1+A^{(1)}_+\alpha_{c1}}\right], \hspace{.5cm}
L_{B1}=\tau^{B}_{c1}.
\ea
While if $B^{(1)}_{+}B^{(1)}_->0$, one can introduce
\be
\alpha_{c1}\equiv e^{2\tau^{B}_{c1}}
 =\half\left[-\left(B^{(1)}_+ +B^{(1)}_-\right)+4r_{c1}^2
  -\sqrt{-4B^{(1)}_{+}B^{(1)}_-+\left(B^{(1)}_+ +B^{(1)}_- -4r_{c1}^2\right)^2}\right],
\ee
and
\ba
\Delta x_{B1} &=& \half\ln\left[\frac{B^{(1)}_+ +\alpha_{c1}}{B^{(1)}_- +\alpha_{c1}}\right]-\half\ln\frac{B^{(1)}_+}{B^{(1)}_-},\\
\Delta t_{B1}&=&\half\ln\left[\frac{A^{(1)}_- +\alpha_{c1}}{A^{(1)}_+ +\alpha_{c1}}\right]-\half\ln\frac{A^{(1)}_-}{A^{(1)}_+}, \\
L_{B1}&=&\tau^{B}_{c1}-\half\ln\left(B^{(1)}_+B^{(1)}_-\right).
\ea
The boundary value of $v$ for the starting point of geodesic is
\be
v_{B1}=t_{B1}+0=\Delta t_{B1}-\textrm{arctanh}\left(\frac{1}{r_{c1}}\right).
\ee

The top BTZ part is similar. At the crossing point $r=r_{c2}$, $\dot{r}_A>0$, and
\be
r_A'=\frac{r_{c2}^2}{p_x}\sqrt{r_{c2}^2+E_A^2-p_x^2}, \hspace{1cm}
E_{B2}=E_A\left(1-\frac{1}{2r_{c2}^2}\right)-\frac{p_x}{2r_{c2}^4}r_A'.
\ee
Note that in general, $E_{B2}\neq E_{B1}$. The value $r_B'\Big|_{c2}$ is important, and determines the behavior just after this crossing point. One finds
\be
r_B'\Big|_{c2}=\left(1-\frac{1}{2r_{c2}^2}\right)r_A'\Big|_{r_{c2}}-\frac{E_A}{2p_x}.
\ee
One has to check the value of  $B^{(2)}_{\pm}\equiv B_{\pm}(p_x,E_{B2})$ first. If $r_B'\Big|_{c2}<0$ and at least one of $B^{(2)}_{\pm}$ is negative, the geodesic will fall to the singularity, otherwise it will eventually goes to the boundary.
Since we are only interested in the ones approaching the boundary, the solution is simplified.
Similarly as above, one introduces a quantity
\be
\alpha_{c2}\equiv e^{2\tau^{B}_{c2}}
 =\half\left[-\left(B^{(2)}_+ +B^{(2)}_-\right)+4r_{c2}^2
  \pm\sqrt{-4B^{(2)}_{+}B^{(2)}_-+\left(B^{(2)}_+ +B^{(2)}_- -4r_{c2}^2\right)^2}\right],
\ee
where ``$\pm$" is the sign of $r_B'\Big|_{c2}$.
The other quantities are
\ba
\Delta x_{B2}=\half\ln\left[\frac{B^{(2)}_+ +\alpha_{c2}}{B^{(2)}_- +\alpha_{c2}}\right], \hspace{.5cm}
\Delta t_{B2}=\half\ln\left[\frac{A^{(2)}_- +\alpha_{c2}}{A^{(2)}_+ +\alpha_{c2}}\right], \hspace{.5cm}
L_{B1}=-\tau^{B}_{c2}.
\ea
The boundary $v$ value is
\be
v_{B2}=\Delta t_{B2}+\textrm{arctanh}\left(\frac{1}{r_{c2}}\right).
\ee

Combining all the three parts together,
\ba
\Delta x &=& \Delta x_A+ \Delta x_{B1}+ \Delta x_{B2}, \\
\Delta t &=& v_{B2}- v_{B1}, \\
L_{reg} &=& L_A + L_{B1} + L_{B2}.
\ea

\subsection{Tests of SSA}

We discuss in previous section about the connection between SSA and the NEC for pure spacelike intervals. In this subsection, the discussion is extended to general spacelike intervals, and we argue the causality somehow preserves the connection. Some examples are checked in Vaidya space, verifying SSA. In the end of the subsection, an example in negative-energy Vaidya space shows that SSA is also violated when the NEC is violated.

\begin{figure}[htbp]
\begin{center}
\includegraphics[scale=.7]{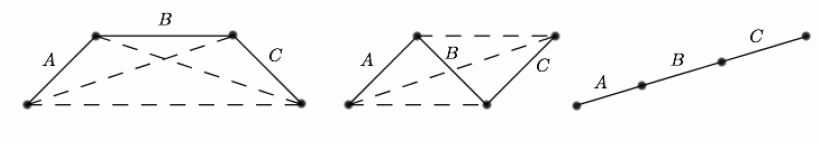}
\caption{The three kinds of boundary intervals to test strong subadditivity: trapezoid, zigzag and linear combinations.
}
\label{fig:3spatialCases}
\end{center}
\end{figure}

For general spacelike intervals, there are basically three kinds of combinations, i.e. trapezoid, zigzag and linear combinations, as in Fig.\ \ref{fig:3spatialCases}.

\begin{figure}[htbp]
\begin{center}
\includegraphics[scale=1.3]{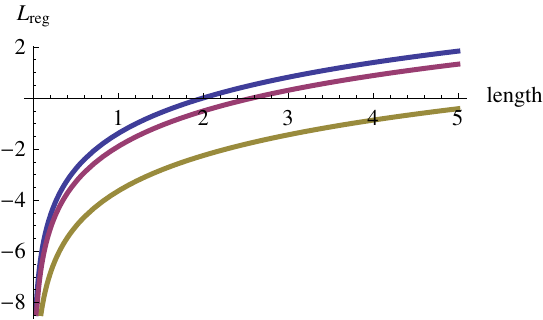}
\includegraphics[scale=1.3]{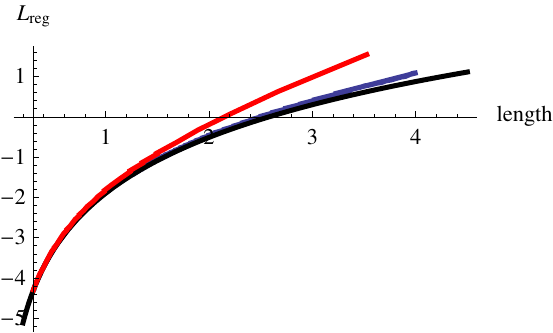}
\caption{$L_{reg}$ as function of length $l$ for straight line intervals on AdS boundary. In the left figure, the bulk is pure AdS space, with different slopes, $\frac{\Delta t}{\Delta x}=\{0,0.5,0.9\}$ from top to bottom respectively. In the right figure, a comparison is given for pure AdS, BTZ and thin-shell Vaidya space. For convenience, we set $\frac{\Delta t}{\Delta x}=0.5$ and fix the center of line on the shell, $v=0$ on boundary. The curves are for BTZ, Vaidya, and pure AdS from top to bottom respectively. All the curves here are obviously concave and monotonically increasing, and thus satisfy strong subadditivity inequalities.
}
\label{fig:Loflength_AdSDiffAngles}
\end{center}
\end{figure}

In all of the three cases, the straight line one is the basic, and we will show some features of this first. As discussed, for a straight line interval, SSA is equivalent to the combination of concavity and monotonically increasing function of $L(l)$, thus we basically need to show the properties of the function. In Fig.\ref{fig:Loflength_AdSDiffAngles}, some comparisons are given. The Vaidya case is much closer to pure AdS than to BTZ, which is reasonable, since although the center of the straight line interval is at $t=0$ on boundary, the interception of the geodesic to the thin-shell has $t>0$, and the geodesic is mainly in AdS bulk. Given the same parameters of the intervals, the corresponding geodesic in BTZ has larger proper length than in AdS, while the one crossing the mass shell is between them. Varying the slope $\Delta t/\Delta x$ of the interval, one finds larger slope gives less proper length. In each given example in the figure, the function $L(l)$ is monotonically increasing and concave, therefore SSA is preserved.

In the following, we test SSA for linear, trapezoid and zigzag combinations of intervals respectively. For simplicity, the lengths are fixed, $l_A=l_C=0.5,~ l_B=1$. The intervals are shifted from AdS to BTZ, thus boundary time value is one of natural parameters. The other variable is the slope of the interval $A$, from which we determine the slopes of $B$ and $C$.

\begin{figure}[htbp]
\begin{center}
\includegraphics[scale=1.1]{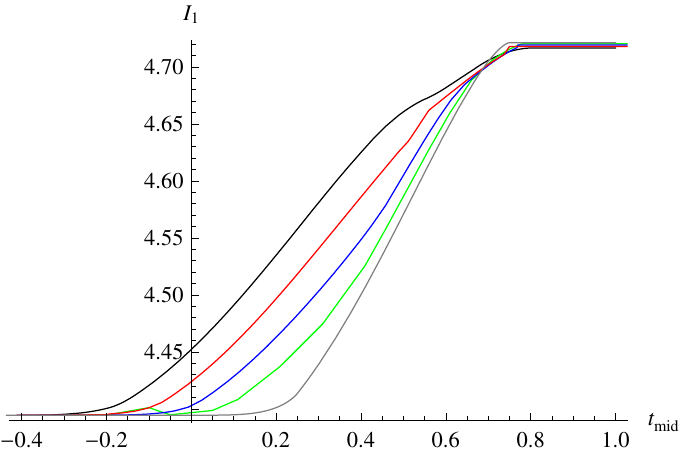}
\includegraphics[scale=1.1]{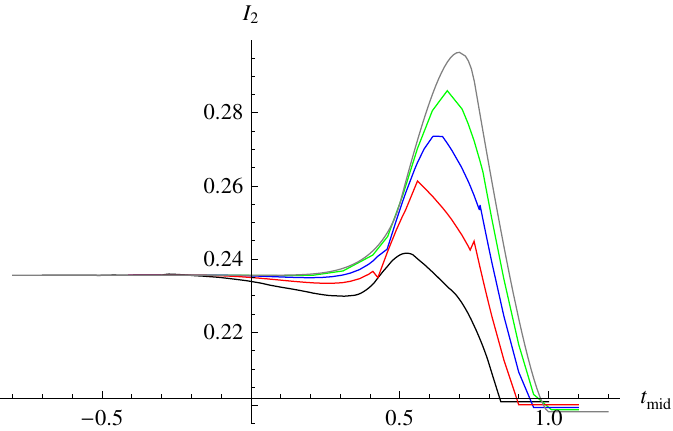}
\caption{SSA is verified with $I_{1,2}$ functions for linear combination of intervals. In the left figure, $\Delta t/\Delta x=\{0,0.204,0.314,0.436,0.577\}$ respectively for the curves from bottom to top. In both figures, the same color corresponds to the same value of $\Delta t/\Delta x$. The length of intervals are set as $l_A=l_C=0.5,l_B=1$. The sharp zigzags are due to numeric errors. The $x$-axis is the boundary time value of interval $B$.
}
\label{fig:VaidyaSSALinearI}
\end{center}
\end{figure}

In Fig.\ref{fig:VaidyaSSALinearI}, SSA is verified for linear combination of intervals. We vary the slope
on $(x,t)$-plane, and find the functions of $I_{1,2}$ are similar to pure spacelike interval case. The $x$-axis indicates the time of the center of interval $B$.(Note that on the boundary, $v=t$.)

\begin{figure}[htbp]
\begin{center}
\includegraphics[scale=1.1]{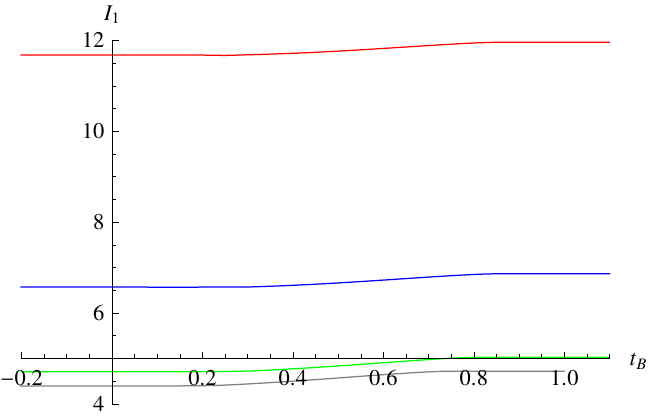}
\includegraphics[scale=1.1]{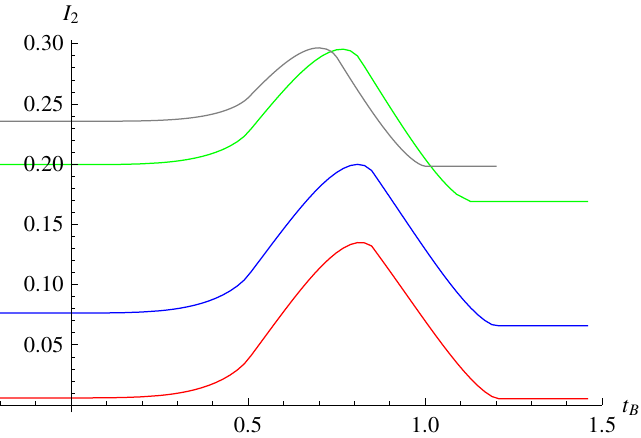}
\caption{SSA is verified with $I_{1,2}$ functions for trapezoid combination of intervals. In the left figure, $\Delta t/\Delta x=\{0,0.314,0.75,0.98\}$ respectively for the curves from bottom to top. In both figures, the same color corresponds to the same value of $\Delta t/\Delta x$. The $x$-axis is the boundary time value of interval $B$.
}
\label{fig:VaidyaSSATrapzoidI}
\end{center}
\end{figure}

For the trapezoid combination, for simplicity, we suppose $A,C$ have the same(with opposite signs)slopes on $(t,x)$-plane, while $B$ is pure spacelike. Several examples of different value of $\Delta t/\Delta x$ are given in Fig.\ref{fig:VaidyaSSATrapzoidI}. As we previously argued, in either pure AdS or BTZ space $I_2=0$ when $A,C$ are lightlike and with opposite signs of $\Delta t/\Delta x$. The vanishing of $I_2$ is approximately confirmed by the red curve which has $\Delta t/\Delta x=0.98$ and intervals $A,C$ are near lightlike. Nevertheless $I_2$ does not vanish around the transition from AdS to BTZ.

\begin{figure}[htbp]
\begin{center}
\includegraphics[scale=1.1]{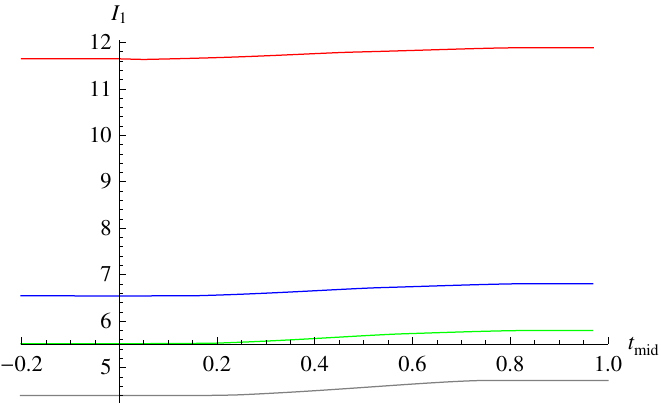}
\includegraphics[scale=1.1]{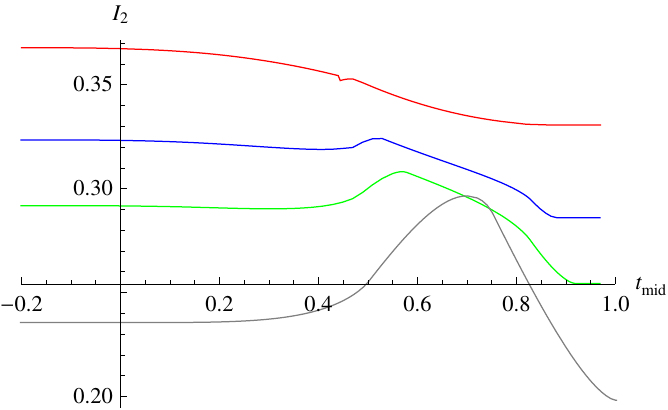}
\caption{SSA is verified with $I_{1,2}$ functions for zigzag combination of intervals. In the left figure, $\Delta t/\Delta x=\{0,0.577,0.75,0.98\}$ respectively for the curves from bottom to top. In both figures, the same color corresponds to the same value of $\Delta t/\Delta x$. The $x$-axis is the boundary time value of the centers of the intervals $A,B$ and $C$.
}
\label{fig:VaidyaSSAZigzagI}
\end{center}
\end{figure}

Now let us check for the zigzag combination. For simplicity, the intervals are intentionally to set that $A$ parallel to $C$, and $AB,BC$ are pure spacelike. Several examples are illustrated in Fig.\ \ref{fig:VaidyaSSAZigzagI}, and for all of these both $I_{1,2}>0$, verifying SSA. We also find numerical evidence that $I_1,I_2$ generally do not vanish even if all the three intervals in zigzag combination are null.

Although in both Fig.\ \ref{fig:VaidyaSSATrapzoidI} and Fig.\ \ref{fig:VaidyaSSAZigzagI}, the $I_1$ functions look pretty flat, a close look shows that they behave more like a smoothed step function, similar to the left figure of Fig.\ref{fig:VaidyaSSALinearI}.

For all the examples examined above, $I_1$ is in general a smoothed step up function, always positive, while $I_2$ has a bump in the transition from AdS to BTZ, and there is no sign to negative value region. Therefore one can conclude that both $I_{1,2}$ are kept positive and SSA is preserved.

\begin{figure}[htbp]
\begin{center}
\includegraphics[scale=1.1]{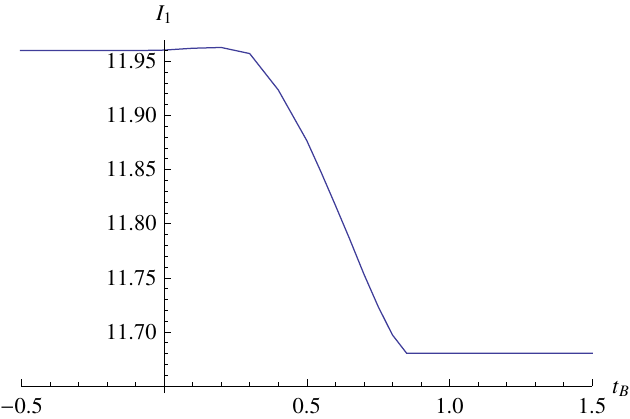}
\includegraphics[scale=1.1]{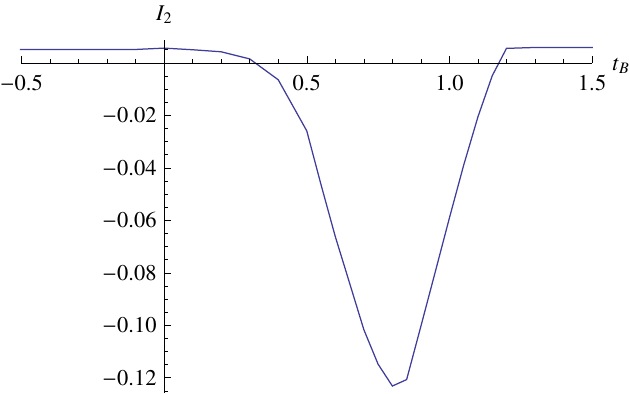}
\caption{The $I_{1,2}$ functions for negative-energy Vaidya, with slope $\Delta t/\Delta x=0.7$ for interval $A$. As usual, $l_A=l_C=0.5, l_B=1$. The $x$-axis is the boundary time value of interval $B$.
}
\label{fig:UnphyVaidyaSSATZI}
\end{center}
\end{figure}

On the other hand, as conjectured, the negative-energy Vaidya space, which violates the NEC, should violate the SSA. An explicit example is shown in Fig.\ref{fig:UnphyVaidyaSSATZI}. Similar to pure spacelike interval on the boundary of negative-energy Vaidya bulk, which has a monotonically increasing but non-concave $L(l_x)$ function(Fig.\ref{fig:UnphysicalVaidyaLoflxlist}), and thus $I_1>0$ while $I_2$ is not always positive.
Compared with Fig.\ref{fig:VaidyaSSATrapzoidI}, for the same combination of intervals on boundary, the $I_{1,2}$ functions seems to exchange the future with the past, and flip the shape upside down, e.g. the bumped $I_2$ in Vaidya v.s. the well-shaped $I_2$ function in negative-energy Vaidya space.

Without time consuming numerical calculation, one can safely say, if the lengths are kept $l_A=l_C=0.5, l_B=1$, the trapezoid combination is the easiest to violate SSA, especially when $A,C$ are near-null. However, the appearance of negative $I_2$ might depend on the length combination one choose.

In addition, for given sets of parameters, $I_{1,2}$ are constant in both pure AdS or BTZ bulk, so as long as these functions start to change, some of the relevant geodesics crosses the thin shell. Since $I_2$ in Fig.\ref{fig:UnphyVaidyaSSATZI} does not jump to negative value but varies smoothly, it suggests us that violation of NEC,  equivalently with $m'(v)<0$ in this example, is not a sufficient condition for SSA's violation. We would like to prove that more formally in our future work.

\acknowledgments

We would like to thank J. McGreevy and E. Tonni for useful conversations. This work was funded by the U.S. Department of Energy under grant DE-FG02-92ER40706 and by the National Science Foundation under CAREER Award PHY-1053842.

\appendix

\section{Negative-energy Vaidya: Details of calculation}\label{appendix:unphysicalVaidya}

Since the constant time geodesics in pure $AdS_3$ or BTZ is discussed already, one needs to focus on the ones crossing the thin shell.
Starting from the symmetric point $(r,v)=(p_x,v_0)$, the geodesic should have constant time behavior in BTZ region,
while $\dot{t}_{AdS}\neq 0$ after the mass shell. Thus it crosses the mass shell at $(r,v)=(r_c,0)$. In BTZ, as given by the standard form,
\be
v=t+\tilde{r}, ~~~\tilde{r}=-\mathrm{arctanh}\left(\frac{1}{r}\right),
\ee
here we have already scaled the mass of the shell $m$ to $1$.
The constant time behavior in BTZ indicates
\be
\tilde{r}(r_c)=-t=-v_0+\tilde{r}(p_x)~~\Longrightarrow~~\frac{(p_x+1)(r_c-1)}{(p_x-1)(r_c+1)}=e^{-2v_0}.
\ee
From the equations of motion (\ref{eqn:EOM_Vaidya_randv}), one can derive some of the quantities relevant. To avoid the Delta function behavior of $v''$, $v'$ has to be continuous, thus
\be
v'(v=0)=2(r'_+-r'_-),
\ee
here the subscript $-$ and $+$ indicate before and after the mass shell, i.e. BTZ and $AdS_3$ respectively.
In BTZ, $E_{BTZ}=0$,
\be
(r'_-)^2=r_c^2(r_c^2-1)(\frac{r_c^2}{p_x^2}-1), \hspace{1cm} v'(v=0)=\frac{r'_-}{r_c^2-1}.
\ee
These together gives
\be
r'_+ ~=~ \left(\frac{1}{2(r_c^2-1)}+1\right)r'_-,
\ee
and $r'_+$ reverses its sign at $r_c^2=\half$.
On the other hand, in AdS,
\be
(r')^2=r^4\left[\left(\frac{r^2}{p_x^2}-1\right)+\frac{E^2}{p_x^2}\right],
\ee
and thus
\be
(r'_+)^2=r_c^4\left[\left(\frac{r_c^2}{p_x^2}-1\right)+\frac{E^2}{p_x^2}\right],
\ee
from which one calculates the conserved momentum along $t$
\be
\label{eqn:Energy_Vaidya_unphy_thinshell}
E= \frac{p_x~r'_-}{2r_c^2(r_c^2-1)} =\pm \frac{1}{2r_c}\sqrt{\frac{r_c^2-p_x^2}{r_c^2-1}}.
\ee

For the geodesics connecting to AdS boundary, the proper length and expansion along $x$-direction are calculated,
\ba
L_{BTZ} &=& 2\ln\left(\sqrt{r^2-1}+\sqrt{r^2-p_x^2}\right)\Bigg|^{r_c}_{p_x}, \\
l_{BTZ} &=& 2\mathrm{ArcTanh}\left(p_x\sqrt{\frac{r^2-1}{r^2-p_x^2}}\right)\Bigg|_{p_x}^{r_c}
  =\ln\left(\frac{\sqrt{r^2-p_x^2}+p_x\sqrt{r^2-1}}{\sqrt{r^2-p_x^2}-p_x\sqrt{r^2-1}}\right)~\Bigg|_{p_x}^{r_c}, \\
L_{AdS} &=& 2\ln\left(r+\sqrt{r^2-p_x^2+E^2}\right) \Bigg|_{r_c}^{\infty}, \\
l_{AdS} &=& \frac{2p_x}{(p_x^2-E^2)r}\sqrt{r^2-p_x^2+E^2} \Bigg|_{r_c}^{\infty}.
\ea

As we discussed, in BTZ there are only two cases for connected symmetric spacelike geodesics with $E=0$, one is totally outside the horizon, and the other totally behind the horizon.
In the first case, the radius of geodesic after crossing the shell continuously grows, and approaches the AdS boundary eventually. The geodesics behind the horizon has complicated behavior, and we will discuss more details below. A close study shows there are four situations, as in Fig.\ \ref{fig:UnphyVaidya_FourCases}:
\begin{figure}[htbp]
\begin{center}
\includegraphics[scale=.8]{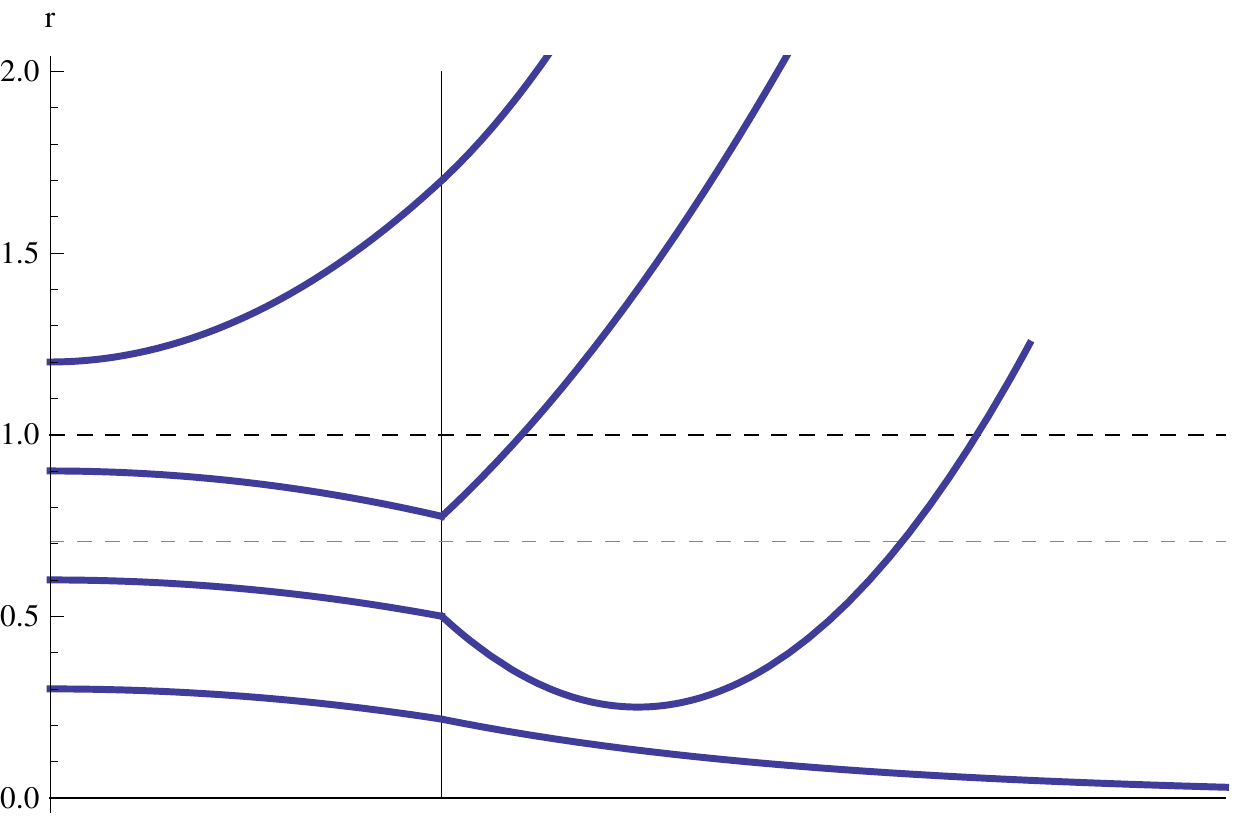}
\caption{The four cases of geodesic in negative-energy thin-shell Vaidya space. The vertical straight line indicates the thin shell at $v=0$, and the horizontal dashed line at $r=1$ is the horizon.  From top to bottom, $r_c>1$, $\half<r_c^2<1$, $r_c^2<\half$ with $p_x^2-E^2>0$, and $r_c^2<\half$ with $p_x^2-E^2<0$ respectively. The geodesics generally encounters a refraction when crossing the shell. The figure is not drawn in scale, only showing the qualitative behavior.
}
\label{fig:UnphyVaidya_FourCases}
\end{center}
\end{figure}

\begin{enumerate}
\item
$r_c>1$, the part in BTZ region is totally outside the horizon.

Obviously, it has some features
\be
r_c>p_x>1, r'_->0\rightarrow r'_+>0
\ee
and
\be
E^2<\frac{1}{4r_c^2}<1<p_x^2<r_c^2.
\ee
With the equation
\be
\dot{r}=\sqrt{r^2+E^2-p_x^2},
\ee
one can solve
\be
r=\frac{1}{2}e^{-\tau}\left(e^{2\tau}+p_x^2-E^2\right),~~\mathrm{or~}\tau=\ln(r+\sqrt{r^2+E^2-p_x^2}).
\ee
At $\tau=\tau_{\infty}\ra\infty, r=r_{\infty}\ra\infty$. The geodesics crosses the shell at
\be
\tau_c=\ln(r_c+\sqrt{r_c^2+E^2-p_x^2})
\ee
Thus the proper length
\be
\tilde{L}=\int d\tau=\tau_{\infty}-\tau_c=\ln(2r_{\infty})-\ln(r+\sqrt{r^2+E^2-p_x^2}).
\ee
Similarly,
\be
x=x_c+\frac{p_x}{r_c(r_c+\sqrt{r_c^2+E^2-p_x^2})}.
\ee
The total proper length and expansion along $x$ are
\ba
L=L_{BTZ}+2\tilde{L}, \hspace{1cm}
l_x=l_{BTZ}+2(x-x_c).
\ea

One simple way to check the transitions is keeping $v_B$ fixed, and change the expansion $l_x$, furthermore, one is able to discuss subadditivity and strong subadditivity properties. However in our method, the initial parameters are $(p_x,v_0)$ of the symmetric point, which could be easily transferred into $(p_x,r_c)$, and it is in general messy to find $(t_b,l_x)$ numerically. It is helpful if we can calculate the constraint with $t_b=t_{boundary}$ fixed. In AdS,
\be
v=t+\tilde{r}=t-\frac{1}{r},\rightarrow ~ v'=t'+\frac{r'}{r^2}=\frac{\dot{t}}{\dot{x}}+\frac{r'}{r^2}=\frac{E}{p_x}+\frac{r'}{r^2},
\ee
and
\ba
v_B &=& 0+\int_{shell}^{boundary}\frac{dv}{dr}dr=\int_{shell}^{boundary}\frac{v'}{r'}dr
= \int_{shell}^{boundary}\left(\frac{1}{r^2}+\frac{1}{r'}\frac{E}{p_x}\right)dr  \nn \\
&=& -\frac{1}{r}\Bigg|_{r_c}^{\infty}+ \frac{E}{p_x}x\Bigg|_{shell}^{\infty}
=\frac{1}{r_c}\frac{E}{r_c(r_c+\sqrt{r_c^2+E^2-p_x^2})}.
\ea
With $t_b$ fixed, one can solve
\be
p_x^2=\frac{r_c^2(4-4r_c t_b+t_b^2)}{(2r_c+t_b-2r_c^2t_b)^2}.
\ee
The constraints from $r_c>p_x>1$ and $t_b>0$ gives that
\be
\label{eqn:vBconstraintrc>1}
\frac{1}{r_c}\leq t_b \leq \frac{4r_c}{4r_c^2-1},,~~\mathrm{and}~~t_b\in(0,\frac{4}{3}).
\ee
For reference, we list some details of the constraints here.
\ba
E\geq 0~&\rightarrow& ~ \frac{1}{r_c}\leq t_b<\frac{2r_c}{2r_c^2-1}, \nn\\
p_x^2\geq 0~&\ra&~ t_b\leq 2(r-\sqrt{r_c^2-1}) \mathrm{~or~} t_b\geq 2(r+\sqrt{r_c^2-1}),
   ~\mathrm{or}~ r_c\leq \frac{4+t_b^2}{4t_b},\nn\\
p_x\geq 1 ~&\ra&~ t_b\leq \frac{4r_c}{4r_c^2-1}.
\ea
All the above constraints combined together gives us the constraint (\ref{eqn:vBconstraintrc>1}).
One can further write down the formula of $(l_x,L)$,
\ba
\label{eqn:LandlxofrcAndvBwithrcgtr1}
l_x &=& -i\pi+2\sqrt{4-4r_c t_b+t_b^2}
  +\ln\left(\frac{2(1-r_c t_b)-\sqrt{4-4r_ct_b+t_b^2}}{2(1-r_c t_b)+\sqrt{4-4r_ct_b+t_b^2}}\right) \nn\\
L &=& \ln\left(\frac{t_b(2r_c+t_b-2r_c^2t_b)^2}{4r_c+t_b-4r_c^2t_b}\right).
\ea
Unfortunately we are not able to find a clean expression of $L(l_x)$, which would be very helpful for the entanglement entropy discussion.

Several geodesics ending on AdS boundary at different value of $t_b$ are given in Fig.\ \ref{fig:UnphysicalVaidya_rcgrt1}. Since they are not very far from $v=0$ at boundary, the curves are pretty close to BTZ curve. For $0<t_b\leq 1$, the $L(l_x)$ curve connects to the bottom AdS curve, which happens when the symmetric point in BTZ moves to $(v=0,r=1)$. However when $t_b>1$, there is a  disconnection, which comes from $E>0$.
\begin{figure}[htbp]
\begin{center}
\includegraphics[scale=.8]{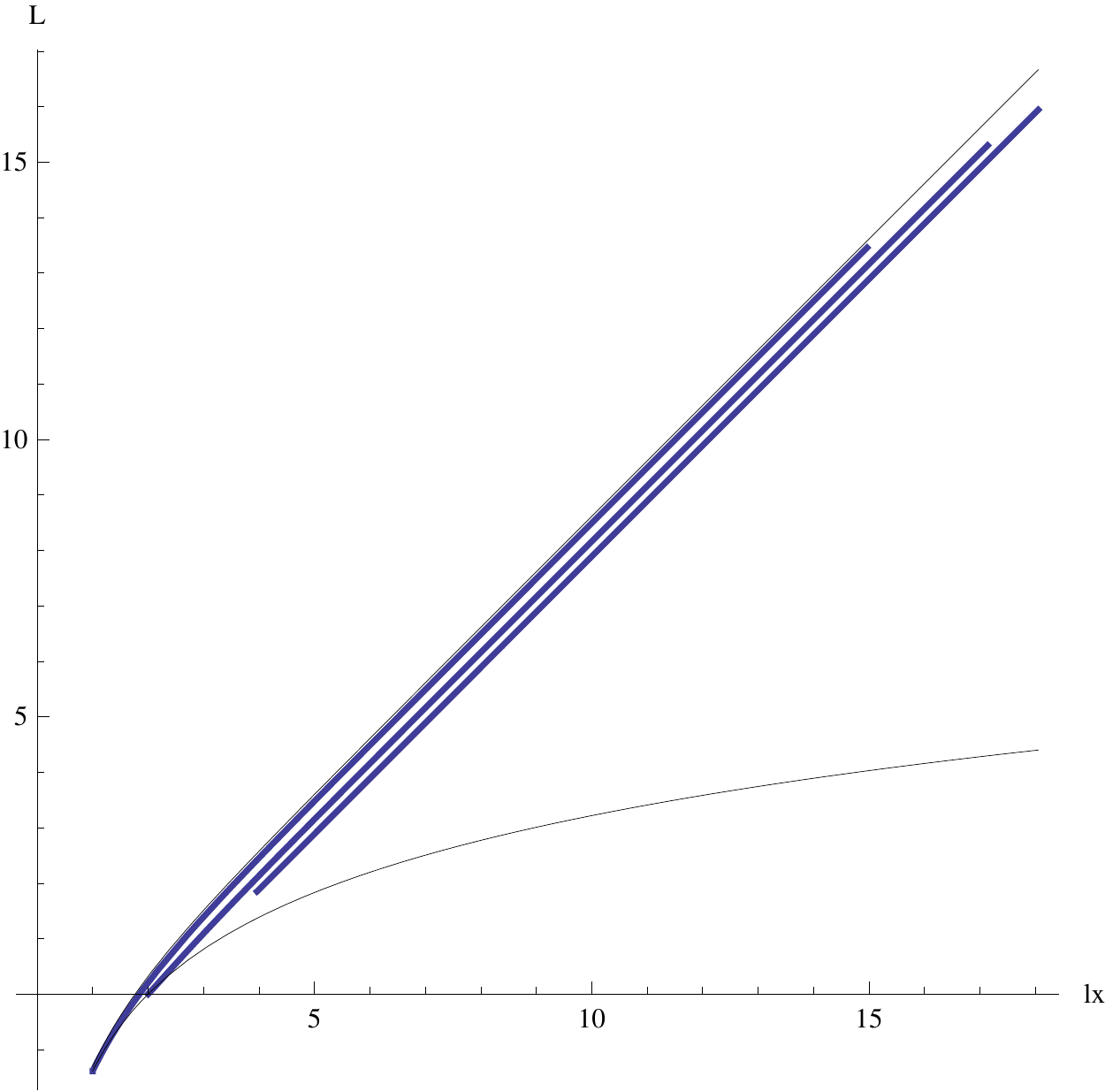}
\caption{Three geodesics ending on AdS boundary at $t_b=(0.5,1,1.3)$ from top to bottom respectively. The top black curve is for BTZ black hole, and the bottom black one is for pure $AdS_3$. The geodesics are pretty close to BTZ curve.
}
\label{fig:UnphysicalVaidya_rcgrt1}
\end{center}
\end{figure}

\item
$\frac{1}{2}<r_c^2<1$, the part in BTZ region is totally behind the horizon, as the second top curve in Fig.\ \ref{fig:UnphyVaidya_FourCases}, with some features:
\be
r_c<p_x<1,~r'_-<0,~r'_+>0,~ E>0.
\ee
Since $r(x)$ is a monotonic increasing function in AdS region, it has
\be
\dot{r}=+\sqrt{r^2+E^2-p_x^2},
\ee
and the solution is
\be
r=\frac{1}{2}e^{-\tau}\left(e^{2\tau}+p_x^2-E^2\right),~~\mathrm{or~}\tau=\ln(r+\sqrt{r^2+E^2-p_x^2}).
\ee
Most functions are the same as the case $r_c>1$, e.g. $\tau_c,\tilde{L},x,t_b,p_x$. Although $\frac{1}{2}< r_c^2<p_x^2\leq 1$ here, different from the first case, the equations leads to the same constraint,
\be
\label{eqn:vBconstraintrc2_Half_1}
\frac{1}{r_c}\leq t_b \leq \frac{4r_c}{4r_c^2-1},~~\mathrm{and}~~t_b\in(1,2\sqrt{2}).
\ee

\item
$0<r_c^2<\frac{1}{2}$ and $p_x^2-E^2>0$. In this case,
\be
r'_-<0,~r'_+<0, E>0,
\ee
and the sign of $\dot{r}$ is changed,
\be
\dot{r}=-\sqrt{r^2+E^2-p_x^2}.
\ee
The solution takes a slightly different form,
\be
r=\frac{1}{2}e^{\tau}\left(e^{-2\tau}+p_x^2-E^2\right),~~\mathrm{or~}\tau=-\ln(r+\sqrt{r^2+E^2-p_x^2}).
\ee
If $p_x^2-E^2>0$, the radius $r$ falls down first, and turns back at $r^2=p_x^2-E^2$, approaching AdS boundary eventually, at $\tau_{\infty}=\ln(2r_{\infty})-\ln(p_x^2-E^2)$. The geodesic crosses the thin shell at $\tau_c=-\ln(r_c+\sqrt{r_c^2+E^2-p_x^2})$.
One can calculate the expansion and proper length
\ba
x&=&x_c+\frac{p_x}{r_c(p_x^2-E^2)}\Big(r_c+\sqrt{r_c^2-p_x^2+E^2}\Big), \\
\tilde{L}&=& \int d\tau=\tau_{\infty}-\tau_c
 =\ln(2r_{\infty})-\ln(p_x^2-E^2)+\ln\Big(r_c+\sqrt{r_c^2-p_x^2+E^2}\Big).  \nn\\
\ea
The constraints on $v_B$ are,
\ba
&\mathrm{when~}r_c< \half,&~~\frac{1}{r_c}<t_b<\frac{3-4r_c^2}{2r_c(1-2r_c^2)}, \nn\\
&\mathrm{when~}r_c=\half, &~~\frac{1}{r_c}<t_b<4, \nn\\
&\mathrm{when~}\half<r_c\leq \frac{1}{\sqrt{2}}, &~~\frac{1}{r_c}<t_b<\frac{2}{r_c},
\ea
and
\be
\label{eqn:vBconstraintrc2_less_Half}
t_b\in(\sqrt{2},\infty).
\ee

\item
$0<r_c^2<\frac{1}{2}$ and $p_x^2-E^2<0$. The geodesic falls to the singularity, in finite proper time
\be
\bar{\tau}=-\half\ln(E^2-p_x^2).
\ee
The proper length is
\be
\tilde{L}=\int d\tau=\bar{\tau}-\tau_c=-\half\ln(E^2-p_x^2)+\ln\Big(r_c+\sqrt{r_c^2+E^2-p_x^2}\Big).
\ee
Although it reaches the singularity in finite proper time, the expansion along $x$ grows to infinity, as well as value of $v$. So we argue it is not necessary to consider any reflection from the singularity $r=0$ and bouncing back to AdS boundary.\footnote{The story is manifestly different in global spacetime, where $x$-direction is compact, and the geodesic will go through the singularity at finite $x$ and end on AdS boundary on "the other" side.}

\end{enumerate}

\bibliographystyle{JHEP}
\bibliography{HolographySSA}

\end{document}